\documentclass[fleqn,usenatbib]{mnras}

\usepackage{newtxtext,newtxmath}

\usepackage{booktabs}
\usepackage{multirow}
\usepackage[normalem]{ulem}
\usepackage{caption}
\usepackage[textwidth=1.5cm,shadow]{todonotes}	
\usepackage{color, soul}	

\usepackage[T1]{fontenc}

\DeclareRobustCommand{\VAN}[3]{#2}
\let\VANthebibliography\thebibliography
\def\thebibliography{\DeclareRobustCommand{\VAN}[3]{##3}\VANthebibliography}

\usepackage{graphicx}	
\usepackage{amsmath}	
 
\usepackage{amssymb}	
\usepackage{hyperref}
\usepackage{xspace}
\usepackage{multirow}
\usepackage{verbatim}

\newcommand{\kms}{\xspace\ensuremath{\rm km\,s^{-1}}}	
\newcommand{\ergcms}{~\ensuremath{\rm erg\,cm^{-2}\,s^{-1}}}

\newcommand{\ergs}{~\ensuremath{\rm erg\,s^{-1}}}

\newcommand{\zsun}{\xspace\ensuremath{{\rm Z_\odot}}\xspace}
\newcommand{\msun}{\xspace\ensuremath{{\rm M_\odot}}\xspace}

\newcommand{\oiii}{[\ion{O}{iii}]\xspace}
\newcommand{\halpha}{H$\alpha$\xspace}
\newcommand{\hbeta}{H$\beta$\xspace}

\newcommand{\nii}{[\ion{N}{ii}]\xspace}
\newcommand{\sii}{[\ion{S}{ii}]\xspace}

\newcommand{\Lbol}{\ensuremath{\rm L_{bol}}\xspace}
\newcommand{\LEdd}{\ensuremath{\rm L_{Edd}}\xspace}
\newcommand{\Mbh}{\xspace\ensuremath{{\rm M_{BH}}}\xspace}
\newcommand{\sigstar}{\xspace\ensuremath{{\sigma_*}}\xspace}
\newcommand{\sigrms}{\xspace\ensuremath{{\sigma^{2}_{rms}\xspace}}}
\newcommand{\Nh}{\ensuremath{{N_{\rm H}}\xspace}}
\newcommand{\G}{\ensuremath{{\Gamma}}\xspace}
\newcommand{\Fx}{\ensuremath{{\rm F_{2-10}}}\xspace}
\newcommand{\A}{\ensuremath{{\rm \AA}}\xspace}
\newcommand{\f}{\xspace\ensuremath{{\to}}\xspace}

\newcommand{\refappendix}[1]{\hyperref[#1]{Appendix~\ref*{#1}}}

\title[Changing Look events in NGC~2992]{The Eddington ratio-dependent `changing look' events in NGC\,2992}

\author[M. Guolo et al.]{Muryel Guolo$^{1,2}$\thanks{E-mail:mguolop1@jh.edu},
    Daniel Ruschel-Dutra$^{1}$,
    Dirk Grupe$^{3}$,
    Bradley M. Peterson$^{4,5,6}$,
    \newauthor
     Thaisa Storchi-Bergmann$^{7}$,
     Jaderson Schimoia$^{8}$,
     Rodrigo Nemmen$^{9}$,
     Andrew Robinson$^{10}$
\\
$^{1}$Departamento de F\'isica - CFM - Universidade Federal de Santa Catarina, 476, 88040-900 Florian\'opolis, SC, Brazil\\
$^{2}$Department of Physics and Astronomy, Johns Hopkins University, 3400 N. Charles St., Baltimore, MD 21218, USA\\
$^{3}$Space Science Center, Morehead State University, 235 Martindale Drive, Morehead, KY 40351, USA\\
$^{4}$ Department of Astronomy, The Ohio State University, 140 W 18th Ave, Columbus, OH 43210, USA\\
$^{5}$ Center for Cosmology and AstroParticle Physics, 191 Woodruff Ave., Columbus, OH 43210, USA\\
$^{6}$ Space Telescope Science Institute, 3700 San Martin Drive, Baltimore, MD 21218, USA\\
$^{7}$ Departamento de Astronomia, Universidade Federal do Rio Grande do Sul. Av. Bento Goncalves 9500, 91501-970 Porto Alegre, RS, Brazil\\
$^{8}$Departamento de F\'isica - CCNE - Universidade Federal de Santa Maria, 97105-90, Santa Maria, RS, Brazil\\
$^{9}$ Instituto de Astronomia, Geof\'isica e Ciências Atmosf\'ericas, Universidade de São Paulo, São Paulo, SP 05508-090, Brazil \\
$^{10}$ School of Physics and Astronomy, Rochester Institute of Technology, 84 Lomb Memorial Drive, Rochester, NY 14623, USA\\
}

\date{Accepted 2021 September 06. Received 2021 August 24; in original form 2020 November 03}

\pubyear{2021}

\begin{document}
\label{firstpage}
\pagerange{\pageref{firstpage}--\pageref{lastpage}}
\maketitle

\begin{abstract}

We present an analysis of historical multi-wavelength emission of the Changing Look (CL) Active Galactic Nucleus (AGN) in NGC 2992, covering epochs ranging from 1978 to 2021, as well as new X-ray and optical spectra.
The galaxy presents multiple Seyfert type transitions from type 2 to intermediate-type, losing and regaining its H$\alpha$ BEL recurrently.
In X-rays, the source shows intrinsic variability with the absorption corrected luminosity varying by a factor of $\sim$ 40.
We rule-out tidal disruption events or variable obscuration as causes of the type transitions, and show that the presence and the flux of the broad H$\alpha$ emission line are directly correlated with the 2-10 keV X-ray
luminosity (L$_{2-10}$): the component disappears at L$_{2-10} \leq 2.6\times10^{42}$\ergcms, this value translates into an Eddington ratio ($\lambda_{\rm Edd}$)
of $\sim$ 1\%.
The $\lambda_{\rm Edd}$ in which the BEL transitions occur is the same as the critical value at which there should be a state transition between a radiatively inefficient accretion flow (RIAF) and a thin accretion disk, such similarity suggests that the AGN is operating at the threshold mass accretion rate between the two accretion modes.
We find a correlation between the narrow Fe K$\alpha$ flux and $\lambda_{\rm Edd}$, and an anti-correlation between full-width at half maximum of H$\alpha$ BEL and $\lambda_{\rm Edd}$, in agreement with theoretical predictions. Two possible scenarios for type transitions are compatible with our results: either the dimming of the AGN continuum, which reduces the supply of ionising photons available to excite the gas in the Broad Line Region (BLR), or the fading of the BLR structure itself occurs as the low accretion rate is not able to sustain the required cloud flow rate in a disk-wind BLR model.

\end{abstract}

\begin{keywords}
galaxies: Seyfert -- galaxies: nuclei -- X-rays: galaxies -- optical: galaxies
\end{keywords}



\section{Introduction}
Supermassive black holes (SMBH) at the centre of galaxies can become active by the accretion of surrounding matter \citep[see][for a review on the feeding mechanisms]{Storchi-Bergmann_2019}. Such active galactic nuclei (AGN) can be identified by features in their optical spectra - such as broad emission lines (BEL) and/or narrow high ionisation emission lines due to the non-stellar ionising continuum - or by their nuclear X-ray emission. Emission lines produced by the gas present in the vicinity of the SMBH are Doppler broadened, forming the broad-line region (BLR), while the narrow-line region (NLR) extends from a few parsecs to hundreds of parsecs beyond the accretion disk. Historically, Type 1 AGNs are those showing both broad (> 10$^{3}$ , < 10$^{4}$\kms) and narrow (< 10$^{3}$\kms) permitted emission lines and narrow forbidden lines in their optical spectra; while Type 2 AGNs lack BELs. Intermediate AGN types were added by \citet{Osterbrock_81}, including types 1.8 and 1.9, classified, respectively, according to the presence of a weak, or absent, \hbeta BEL, while retaining an \halpha BEL. 

AGN unification models attribute differences in AGN type to the viewing angle toward an axisymmetric, parsec scale obscuring structure, which can block our direct view of the AGN central engine \citep{Antonucci_93,Urry_95} and is responsible for the lack of BEL in Type 2 AGNs. However, a relatively small number of AGN show transitions from one type to another in a few years; these objects are known as ``changing look'' (CL) AGNs. CL-AGNs have changed the widely accepted AGN paradigm, not only in the orientation-based unified model but also standard accretion disk models \citep[see][for a review of the so-called Quasar Viscosity Crisis]{Lawrence_18}. Their extreme and fast variability has motivated alternatives and/or modifications for both the AGN Unified Model \citep[e.g.][]{Nicastro_2000,Elitzur_2006,Elitzur_2014} and for the standard \citet{Shakura_73} accretion disk model \citep[e.g.][]{Sniegowska_2020,Jiang_2020}

Until recently there were only a handful of CL events reported, including NGC~3516 \citep{Andrillat_71}, NGC 1566 \citep{Pastoriza_70,Oknyansky_2019}, Mrk 590 \citep{Denney_2014}, NGC 2617 \citep{Shappee_14}, NGC 7603 \citep{Tohline_76}, Mrk 1018 \citep{Cohen_86}, NGC 1097 \citep{Storchi-Bergmann_93}, NGC 3065 \citep{Eracleous_01} and NGC 7582 \citep{Aretxaga_99}. In 2015 \citeauthor{Lamassa_15} discovered the first CL Quasar and since then the number of known CL-AGNs has increased rapidly 
\citep[e.g.][]{MacLeod_2016,McElroy_16,Runnoe_16,Yang_18,Ruan_2016,Sheng_2017,Kollatschny_2018,Guo_19,Wang_2019,Graham_2019,Kollatschny_2020,Graham_2020}; however, the number is still fewer than a hundred. Although rare, these objects provide valuable constraints for our models of the central engine.

Some of these CL might be successfully explained via variable obscuration \citep[e.g.,][]{Matt_2003,Puccetti_07,Risaliti_2010,Marinucci_2013,Marchese_2012}: in models where the obscuring material has a patchy distribution \citep[e.g.][]{Nenkova_2008, Elitzur_2012}, the dynamical movement of the dust clouds could result in a change of classification. 
Transient events such as tidal disruption events (TDE) of a star by the SMBH were also claimed as possible drivers of CL events \citep{Eracleous_1995}; however, this scenario does not apply to sources presenting multiple (and quasi-periodic) CL events.

In a third scenario, it has been argued that changes in type are expected with variations in accretion rate. In this scenario either the BELs disappear (reappear) due to the reduction (increase) of ionising photons available to excite the gas in the vicinity of the black hole \citep[e.g.][]{Storchi-Bergmann_2003,Lamassa_15} or the BLR itself disappears, given that in disk-wind BLR models \citep[e.g.][]{Nicastro_2000,Elitzur_2006,Elitzur_2009} the broad line emission region follows an evolutionary sequence that is directly related to the accretion rate of the compact source. In such disk-wind BLR models, a low accretion efficiency is not able to sustain the required cloud flow rate responsible for the formation of the BLR clouds which gives origin to the distinct (intrinsic) AGN types, as proposed by \citet{Elitzur_2014}.

NGC\,2992 is a highly inclined, $i\sim70^\circ$,  nearby \citep[z = 0.00771,][]{Keel_96} Seyfert galaxy. The galaxy was the subject of several studies and target of all major X-ray missions due to its variability. In the X-rays, it steadily declined in observed flux from 1978, when it was observed by
\textit{HEAO1} \citep{Mushotzky_82} at a flux level of about $8\times10^{-11}$\ergcms
until 1994 when it was observed by \textit{ASCA}
\citep{Weaver_96} at a fainter flux level by about a factor 20. Then it underwent a rapid recovery: in
1997 it was observed by \textit{BeppoSAX} at a flux level somewhat higher than in 1994, while in 1998 the source fully recovered its \textit{HEAO1} brightness \citep{Gilli_00}. In 2003, when observed by \textit{XMM-Newton}, the flux was even higher,
about $10^{-10}$\ergcms \citep{Shu_10}. The source
was then observed by \textit{Suzaku} on November/December 2005, and found in a much fainter state, almost an order of magnitude fainter than in the \textit{XMM-Newton} observation. 

The Rossi X-ray Timing Explorer \citep[RXTE,][]{Swank_2006} monitoring campaign  between early March 2005 and late January
2006 found large amplitude (almost an order of magnitude) variability, indicating that apart from the long-term variability of the source, it also presents short-term outbursts with variations in the order
of days, while no significant variation of the primary power-law index was found \citep{Murphy_07};  at the end of the campaign the source was again at a low flux state, confirmed by \textit{Suzaku}, at a flux level of $\sim 1\times10^{-11}$\ergcms \citep{Ptak_07,Yaqoob_2007} and \textit{Chandra}, at flux level of 
$\sim0.3\times10^{-11}$\ergcms \citep[its historical minimum flux,][]{Murphy_17}. In 2010, NGC 2992 was observed eight times for $\sim$ 40\,ks by
\textit{XMM-Newton} with a 2–10 keV flux ranging from $0.3\times10^{-11}$\ergcms to $1.3\times10^{-11}$\ergcms. A further 2013 \textit{XMM-Newton} observation caught the source in a higher but still low flux state
($1.6\times10^{-11}$\ergcms). In 2015, observations performed simultaneously with
\textit{Swift} and \textit{NuSTAR} found a 2-10 keV flux of $6\times10^{-11}$\ergcms \citep{Marinucci_18}. The most recent published observations were performed again with \textit{XMM-Newton} in May 2019, when the flux was again at high levels, ranging from $6.8$ to  $10\times10^{-11}$\ergcms \citep{Marinucci_20}.

The optical variability of the source was first explored by \citet{Gilli_00} who noticed that in 1999 the \halpha BEL was present again, as it was in early observations of the galaxy by \citet{Veron_80} at the end of the 1970s which led to its first classification as a Seyfert 1.9, in contrast to the Type 2 spectra seen in 1994 by \citet{Allen_99}. Later, \citet{Trippe_08} showed that from the beginning of 2006 to the middle of 2007 the source had lost its BEL. Recently \citet{Schnorr-Muller_16} and \citet{Guolo_Pereira_2021} not only showed that the \halpha BEL was detectable again but also, for the first time, a very faint \hbeta BEL was claimed to be present.

Besides being a very well studied galaxy, to the best of our knowledge no study has yet explored the entire X-ray and optical historical emission of NGC~2992, nor attempted to link the behaviour in the two wavelength bands. These are our goals in this paper. The paper is organised as follows: in \autoref{sec:data} we describe the data collection. The analysis of the data is described in  \autoref{sec:analyses}, while the results are discussed in \autoref{sec:discussion}. Our conclusions appear in \autoref{sec:conclusion}.

\section{Data Collection}
\label{sec:data}

\begin{figure*}
    \centering
    \includegraphics[width=0.9\textwidth]{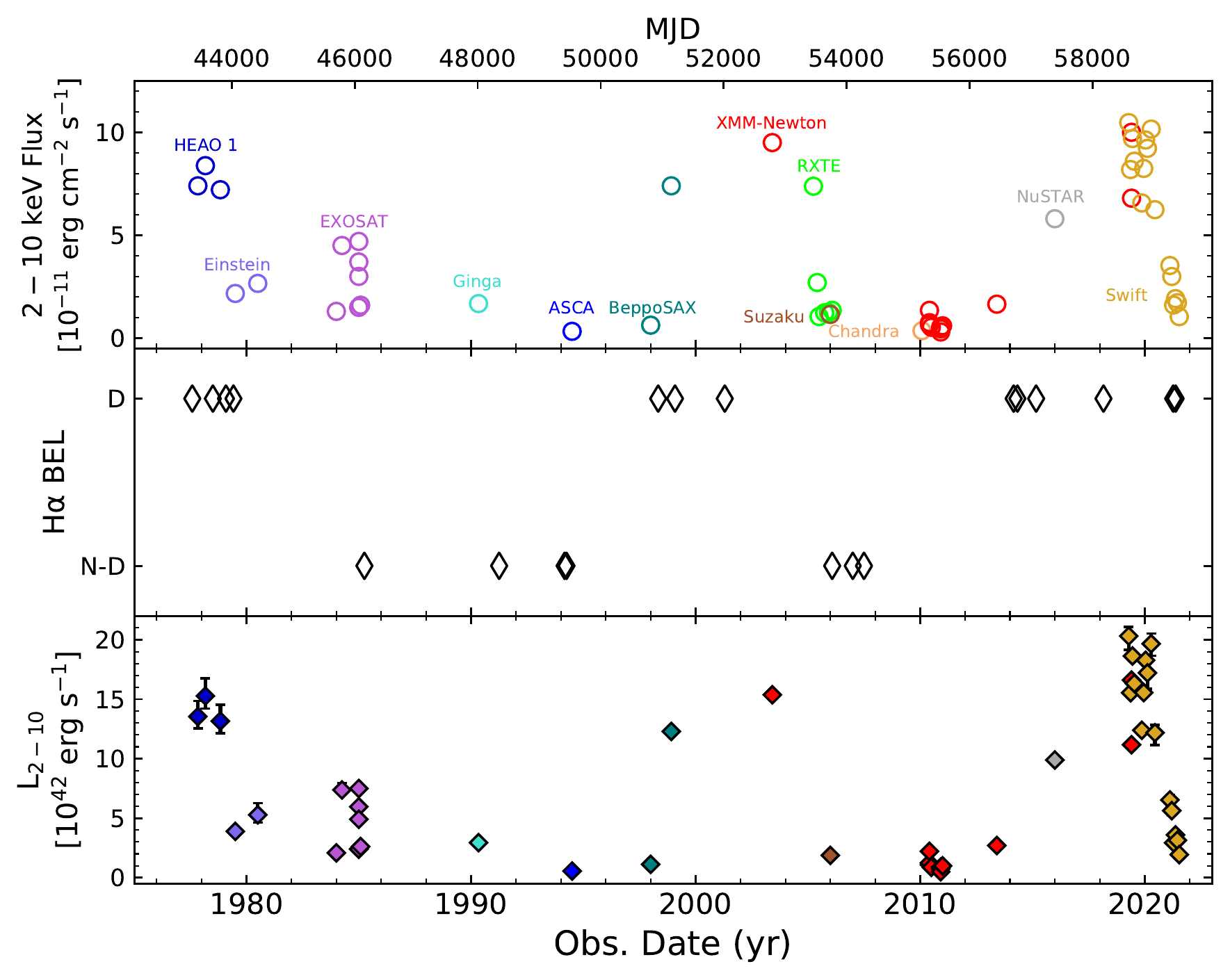}
    \caption{Top: Historical 2-10 keV light curve of the source. Middle: Historical detections (D) and non-detections (N-D) of the broad \halpha emission line. Bottom: Intrinsic (absorption corrected) 2-10 keV Luminosity (L$_{\rm 2-10}$) light curve. Each colour represents a distinct X-ray mission, and will be maintained throughout the entire paper. In the L$_{\rm 2-10}$ panel, the points without error bars have errors smaller then 3$\times$10$^{41}$ erg s$^{-1}$, see \autoref{table:xray} for the exact values.}
    \label{fig:L_x_time}
\end{figure*}

This paper results from a careful re-examination of the entire literature on NGC~2992, both X-rays and optical spectra, from its discovery as an X-ray source \citep{Cooke_78,Ward_80} and classification as a Seyfert galaxy \citep{Osmer_74} to the most recent optical and X-ray spectra published to date \citep[][]{Marinucci_20,Guolo_Pereira_2021}.

\subsection{Historical X-ray Data}

The X-ray spectrum of NGC~2992 is
very well fitted by an absorbed power-law which accounts for most of the X-ray emission. However, the spectrum also presents a narrow iron K$\alpha$ component at 6.4 kev,
which becomes more intense at high flux levels \citep[][see also \autoref{sec:Xray_var}]{Yaqoob_2007,Marinucci_18,Marinucci_20},
We collected the fluxes in the 2-10~keV X-ray band (F$_{\rm 2-10}$) and the best-fitting parameters for the absorbed continuum model, namely the power-law continuum photon indexes ($\Gamma$)\footnote{The power law index in the continuum model: $F(E) \propto E^{-\Gamma}$.} and the absorbed material column densities ($N_{\rm H}$), as reported by the authors throughout the literature on the source. For the more recent spectra, where the narrow Fe K$\alpha$ is detected we also collected their measured fluxes, F(Fe K$\alpha$). In the top panel of \autoref{fig:L_x_time}, we show the historical 2-10 keV light curve. In \autoref{table:xray}, we show the best-fitting continuum model parameters, the Fe K$\alpha$ fluxes, the observation dates, and references. The $N_{\rm H}$ are only accounting for the absorption at the redshift of the source therefore not including Galactic absorption, $N_{\rm H \ Gal} \approx$ 5.5$\times$10$^{20}$ cm$^{-2}$ \citep[][]{Kalberla_2005}. Errors in the X-rays parameters correspond to the 90\% confidence level for the parameter of interest ($\Delta \chi^2 = 2.7$), as reported by the authors. For the papers in which the authors reported their uncertainties in another scale, e.g., $\Delta  \chi^2 = 4.6$, the errors were re-scaled using a Python package (\href{https://github.com/muryelgp/asymmetric\_uncertainties}{\tt asymmetric\_uncertainties}) for treatment of asymmetric statistical errors derived from confidence levels, which employs a combination of the methods proposed by \citet{Barlow_04} and \citet{Possolo_19}. The method and a brief discussion on uncertainties can be found in Appendix \ref{app1}. The propagation of the uncertainties for the X-ray derived quantities (see section 3.1 and \autoref{table:xray}) is also performed with this same method, see appendix \ref{app1}.

\subsection{New Swift X-ray Data}

The X-ray Telescope \citep[XRT,][]{Burrows_2005} on-board of the \textit{Neil Gehrels Swift} Observatory started its first monitoring campaign of NGC~2992 in March 2019, continuing it throughout the entire year (except for the period of early-August to late-September when the galaxy is in sun-constrained).
During 2020 few visits were made, while a second monitoring campaign started in January 2021; this campaign is still ongoing at the time of this publication. However, in this work, we will limit ourselves to all data taken before June 15th 2021.
All observations were performed in the Photon Counting mode \citep[PC mode][]{Hill_2004}, with exposure times varying between 1 and 2~ks, at a cadence varying between 5 and 7 days. Given that in this work we are only interested in the medium to long-term variability (years/months), not in the short-term (days/weeks) variability, the observations were combined monthly with each resulting spectrum representing the average behaviour of the AGN at the given month. The data in the original cadence may be explored in a future publication by the P.I. of the monitoring campaigns (Middei et al., in prep). Data were reduced with the task \textit{xrtpipeline} version 0.13.5. As a very nearby source in PC mode, pileup effects may be present, therefore the extractions were made with an annulus aperture that excludes the inner 5 pixels of the Point Spread Function, using \textit{XSELECT} version 2.4x.
The spectral data were re-binned with at least 20 photons per bin using \textit{grppha} version 3.0.0. The auxiliary response files were created with \textit{xrtmkarf} and corrected using the exposure maps, and the standard response matrices \textit{swxpc0to12s6\_20130101v014.rmf}.
The 0.3-10.0 keV spectra were analysed with \textit{pyXSPEC} version 2.0.3 \citep{Arnaud_1996} by fitting an absorbed power-law model, using again a Galactic absorption, $N_{\rm H \ Gal} \approx$ 5.5$\times$10$^{20}$ cm$^{-2}$ \citep[][]{Kalberla_2005}. Best-fitting F$_{\rm 2-10}$, $N_{\rm H}$, $\Gamma$ and their 90\% confidence level error are presented in \autoref{table:xray}, while the flux is plotted on the top panel of \autoref{fig:L_x_time}.

The new \textit{Swift} data shows that the AGN was at its high state during entire 2019 year as well as in the 2020, reaching its maximum observed flux (10.4$\times $10$^{-11}$\ergcms) in March 2019. However early 2021 data shows the AGN constantly declining its flux, since January (3$\times10^{-11}$\ergcms), with March-May observations having a flux of 1.5-2$\times10^{-11}$\ergcms comparable with 2013 \textit{XMM-Newton} data \citep{Marinucci_18}, while June data reaches an even lower flux value of 1$\times10^{-11}$\ergcms, showing that the AGN is again transitioning to its low flux state. 

\begin{table*}
\caption{Historical X-ray data for NGC~2992.}
\centering
\begin{tabular}{|c|c|c|c|c|c|c|c|} 
\toprule
Obs. Date$^{(\rm a)}$    & Satellite                   & F$_{\rm 2-10}^{(\rm b)}$   & $N_{\rm H}^{(\rm c)}$            & $\Gamma$        & F(Fe K$\alpha$)$^{(\rm d)}$                                      & Reference                                                & L$_{\rm 2-10}^{(\rm e)}$                   \\ 
\midrule
10/1977  & \multirow{3}{*}{HEAO 1}     & 7.40        & \multirow{3}{*}{16.0$_{-6.0}^{+8.0}$ } & \multirow{3}{*}{1.79$_{-0.07}^{+0.09}$ } & – & \multirow{3}{*}{\citet{Mushotzky_82} } & 13.48$_{-0.99}^{+1.40}$   \\
~02/1978 &                             & 8.38        &                                        &                                          & – &                                                          & 15.30$_{-1.13}^{+1.52}$    \\
10/1978  &                             & 7.21        &                                        &                                          & – &                                                          & 13.09$_{-0.85}^{+1.34}$    \\ 
\midrule
06/1979  & \multirow{2}{*}{Einstein}   & 2.17        & 14.3$_{-4.0}^{+6.0}$                   & 1.82$_{-0.14}^{+0.14}$                  & –  & \multirow{2}{*}{\citet{Maccacaro_82}}  & 3.89$_{-0.20}^{+0.30}$     \\
06/1980  &                             & 2.66        & 22.3$_{-10.7}^{+15.2}$                 & 1.92$_{-0.47}^{+0.47}$                  & –  &                                                          & 5.25$_{-0.64}^{+1.01}$      \\ 
\midrule
12/1983  & \multirow{8}{*}{EXOSAT}     & 1.30        & 7.0$_{-1.2}^{+3.5}$                    & 1.46$_{-0.16}^{+0.23}$                  & –  & \multirow{8}{*}{\citet{Turner_89}}     & 2.09$_{-0.03}^{+0.10}$     \\
03/1984   &                             & 4.50        & 7.8$_{-1.6}^{+6.2}$                    & 1.68$_{-0.22}^{+0.31}$                  & –  &                                                          & 7.37$_{-0.17}^{+0.59}$    \\
12/1984  &                             & 4.70        & 6.0$_{-0.6}^{+1.7}$                    & 1.64$_{-0.09}^{+0.07}$                   & – &                                                          & 7.50$_{-0.07}^{+0.18}$   \\
12/1984  &                             & 3.70        & 7.0$_{-0.3}^{+1.2}$                    & 1.56$_{-0.06}^{+0.10}$                 & –   &                                                          & 5.96$_{-0.03}^{+0.12}$    \\
12/1984  &                             & 3.00        & 8.6$_{-0.3}^{+2.6}$                    & 1.48$_{-0.11}^{+0.15}$                 & –   &                                                          & 4.91$_{-0.03}^{+0.27}$     \\
12/1984  &                             & 1.50        & 6.9$_{-1.8}^{+7.1}$                    & 1.49$_{-0.27}^{+0.46}$                 & –   &                                                          & 2.42$_{-0.07}^{+0.23}$     \\
12/1984  &                             & 1.50        & 5.6$_{-1.2}^{+5.2}$                    & 1.58$_{-0.20}^{+0.39}$                 & –   &                                                          & 2.39$_{-0.05}^{+0.19}$      \\
01/1985  &                             & 1.60        & 7.4$_{-1.2}^{+4.7}$                    & 1.67$_{-0.17}^{+0.31}$                 & –   &                                                          & 2.61$_{-0.05}^{+0.18}$      \\ 
\midrule
04/1990  & GINGA                       & 1.68        & 13.6$_{-3.3}^{+3.3}$                   & 1.64$_{-0.08}^{+0.08}$                 & –   & \citet{Nandra_94}                      & 2.94$_{-0.12}^{+0.12}$     \\ 
\midrule
06/1994  & ASCA                        & 0.33        & 10.1$_{-2.3}^{+2.7}$                   & 1.70$^{(\rm f)}$                     & –                  & \citet{Weaver_96}                     & 0.56$_{-0.02}^{+0.02}$      \\ 
\midrule
12/1997  & \multirow{2}{*}{BeppoSAX}   & 0.63        & 14.0$_{-4.0}^{+5.0}$                   & 1.72$_{-0.12}^{+0.13}$                & –    & \multirow{2}{*}{\citet{Gilli_00}}      & 1.12$_{-0.06}^{+0.07}$   \\
11/1998  &                             & 7.40        & 9.0$_{-0.3}^{+0.3}$                    & 1.70$_{-0.02}^{+0.02}$               & –     &                                                          & 12.29$_{-0.05}^{+0.05}$   \\ 
\midrule
05/2003  & XMM-Newton                  & 9.50        & 6.5$_{-0.2}^{+0.3}$                    & 1.83$_{-0.04}^{+0.06}$                & –    & \citet{Shu_10}                         & 15.37$_{-0.05}^{+0.07}$    \\ 
\midrule
03/2005  & \multirow{6}{*}{RXTE}       & 7.38        & –${(\rm g)}$                                     & 1.71$_{-0.03}^{+0.03}$         & –           & \multirow{6}{*}{\citet{Murphy_07}}     & –                                            \\
05/2005  &                             & 2.70        & –                                      & 1.72$_{-0.04}^{+0.04}$                & –    &                                                          & –                                              \\
06/2005  &                             & 1.05        & –                                      & 1.75$_{-0.10}^{+0.10}$                & –    &                                                          & –                                             \\
09/2005  &                             & 1.23        & –                                      & 1.85$_{-0.12}^{+0.13}$                 & –   &                                                          & –                                             \\
11/2005  &                             & 1.25        & –                                      & 1.96$_{-0.11}^{+0.12}$                & –    &                                                          & –                                              \\
01/2006  &                             & 1.35        & –                                      & 1.76$_{-0.08}^{+0.08}$                & –    &                                                          & –                                              \\ 
\midrule
12/2005  & SUZAKU                      & 1.15        & 8.0$_{-0.4}^{+0.6}$                    & 1.57$_{-0.03}^{+0.05}$               & 3.99$^{+0.80}_{-0.80}$    & \citet{Yaqoob_2007}                        & 1.88$_{-0.01}^{+0.01}$    \\ 
\midrule
01/2010  & CHANDRA                     & 0.36        & –                                      & –                                      & –  & \citet{Murphy_17}                      & –                                               \\ 
\midrule
05/2010  & \multirow{9}{*}{XMM-Newton} & 0.65        & 9.2$_{-1.0}^{+1.0}$                    & 1.67$_{-0.05}^{+0.05}$               & 3.51$^{+0.35}_{-0.35}$    & \multirow{10}{*}{\citet{Marinucci_18}} & 1.08$_{-0.01}^{+0.01}$    \\
05/2010  &                             & 0.75        & 8.6$_{-1.0}^{+1.0}$                    & 1.64$_{-0.04}^{+0.04}$                & 2.96$^{+0.42}_{-0.42}$    &                                                          & 1.24$_{-0.01}^{+0.02}$    \\
05/2010  &                             & 1.35        & 8.4$_{-1.0}^{+1.0}$                    & 1.61$_{-0.04}^{+0.04}$                & 4.49$^{+0.48}_{-0.48}$    &                                                          & 2.22$_{-0.03}^{+0.03}$    \\
06/2010  &                             & 0.53        & 8.5$_{-1.0}^{+1.0}$                    & 1.67$_{-0.05}^{+0.05}$                & 3.69$^{+0.32}_{-0.32}$   &                                                          & 0.87$_{-0.01}^{+0.01}$     \\
11/2010  &                             & 0.54        & 8.0$_{-1.0}^{+1.0}$                    & 1.67$_{-0.05}^{+0.05}$                 & 3.64$^{+0.34}_{-0.34}$   &                                                          & 0.88$_{-0.01}^{+0.01}$     \\
11/2010  &                             & 0.45        & 8.0$_{-1.0}^{+1.0}$                    & 1.70$_{-0.06}^{+0.06}$                 & 3.70$^{+0.35}_{-0.35}$   &                                                          & 0.74$_{-0.01}^{+0.01}$     \\
11/2010  &                             & 0.30        & 8.1$_{-1.0}^{+1.0}$                    & 1.71$_{-0.09}^{+0.09}$                 & 3.19$^{+0.32}_{-0.32}$   &                                                          & 0.49$_{-0.01}^{+0.01}$     \\
12/2010  &                             & 0.60        & 9.0$_{-1.0}^{+1.0}$                    & 1.68$_{-0.04}^{+0.04}$                 & 3.00$^{+0.37}_{-0.37}$  &                                                          & 1.00$_{-0.01}^{+0.01}$     \\
05/2013  &                             & 1.65        & 8.1$_{-1.0}^{+1.0}$                    & 1.63$_{-0.06}^{+0.06}$                 & 5.66$^{+1.12}_{-1.12}$  &                                                          & 2.70$_{-0.03}^{+0.03}$    \\
12/2015  & NuSTAR                      & 5.80        & 11.0$_{-2.0}^{+2.0}$                   & 1.72$_{-0.03}^{+0.03}$                  &8.68$^{+1.92}_{-1.92}$  &                                                                & 9.89$_{-0.24}^{+0.26}$     \\ 
\midrule
05/2019  & \multirow{2}{*}{XMM-Newton} & 10.00       & 9.0$_{-1.6}^{+1.6}$                    & 1.68$_{-0.10}^{+0.10}$                &12.96$^{+1.60}_{-1.60}$   & \multirow{2}{*}{\citet{Marinucci_20}}  & 16.59$_{-0.34}^{+0.34}$    \\
05/2019  &                             & 6.80        & 8.5$_{-1.6}^{+1.6}$                    & 1.63$_{-0.06}^{+0.06}$                 &12.96$^{+1.60}_{-1.60}$  &                                                          & 11.18$_{-0.22}^{+0.23}$    \\
\midrule

03/2019 &  \multirow{16}{*}{Swift}        &  10.47$_{-0.50}^{+0.38}$ &   9.3$_{-1.4}^{+1.6}$ &  1.77$_{-0.15}^{+0.16}$ &    – &   \multirow{16}{*}{This Work}    &  20.32$_{-1.16}^{+0.77}$ \\
04/2019 &          &   8.19$_{-0.20}^{+0.16}$ &   7.9$_{-0.7}^{+0.8}$ &  1.50$_{-0.08}^{+0.08}$ &    – &     &  15.55$_{-0.38}^{+0.30}$ \\
05/2019 &          &   9.70$_{-0.12}^{+0.12}$ &   8.9$_{-0.5}^{+0.5}$ &  1.63$_{-0.05}^{+0.05}$ &    – &     &  18.64$_{-0.19}^{+0.21}$ \\
06/2019 &          &   8.60$_{-0.20}^{+0.15}$ &   7.9$_{-0.7}^{+0.7}$ &  1.51$_{-0.07}^{+0.07}$ &    – &     &  16.33$_{-0.36}^{+0.31}$ \\
10/2019 &          &   6.57$_{-0.22}^{+0.13}$ &   7.3$_{-0.9}^{+1.0}$ &  1.45$_{-0.09}^{+0.09}$ &    – &     &  12.39$_{-0.37}^{+0.31}$ \\
11/2019 &          &   8.24$_{-0.17}^{+0.14}$ &   7.4$_{-0.5}^{+0.5}$ &  1.43$_{-0.06}^{+0.06}$ &    – &     &  15.56$_{-0.28}^{+0.28}$ \\
12/2019 &          &   9.63$_{-0.24}^{+0.20}$ &   7.8$_{-0.6}^{+0.6}$ &  1.53$_{-0.06}^{+0.07}$ &    – &     &  18.28$_{-0.43}^{+0.30}$ \\
01/2020 &          &   9.22$_{-0.57}^{+0.72}$ &   5.7$_{-1.9}^{+2.2}$ &  1.49$_{-0.22}^{+0.23}$ &    – &     &  17.22$_{-1.34}^{+0.82}$ \\
03/2020 &          &  10.16$_{-0.60}^{+0.40}$ &   9.6$_{-1.9}^{+2.2}$ &  1.69$_{-0.18}^{+0.20}$ &    – &     &  19.67$_{-1.01}^{+0.87}$ \\
05/2020 &          &   6.24$_{-0.57}^{+0.38}$ &  10.7$_{-2.7}^{+3.1}$ &  1.68$_{-0.27}^{+0.29}$ &    – &     &  12.18$_{-1.04}^{+0.66}$ \\
01/2021 &          &   3.53$_{-0.27}^{+0.19}$ &   5.5$_{-1.8}^{+2.1}$ &  1.24$_{-0.21}^{+0.22}$ &    – &     &   6.54$_{-0.58}^{+0.41}$ \\
02/2021 &          &   3.00$_{-0.17}^{+0.12}$ &   7.0$_{-1.4}^{+1.6}$ &  1.40$_{-0.15}^{+0.16}$ &    – &     &   5.63$_{-0.39}^{+0.20}$ \\
03/2021 &          &   1.61$_{-0.12}^{+0.10}$ &   3.3$_{-1.2}^{+1.6}$ &  1.11$_{-0.17}^{+0.19}$ &    – &     &   2.93$_{-0.20}^{+0.14}$ \\
04/2021 &          &   1.92$_{-0.11}^{+0.09}$ &   6.2$_{-1.4}^{+1.6}$ &  1.44$_{-0.17}^{+0.18}$ &    – &     &   3.59$_{-0.18}^{+0.17}$ \\
05/2021 &          &   1.72$_{-0.13}^{+0.10}$ &   3.6$_{-1.3}^{+1.7}$ &  1.10$_{-0.17}^{+0.19}$ &    – &     &   3.14$_{-0.22}^{+0.19}$ \\
06/2021 &          &   1.04$_{-0.21}^{+0.08}$ &   5.2$_{-3.7}^{+4.4}$ &  1.44$_{-0.46}^{+0.50}$ &    – &     &   1.94$_{-0.37}^{+0.14}$ \\

\bottomrule

\end{tabular}
\label{table:xray}

\footnotesize{(a) mm/yyyy. (b) in units of {$\rm 10^{-11} \ erg \ cm^{-2} \ s^{-1}$}. (c) in units of $\rm 10^{21} \ cm^{-2}$. (d) in units of $10^{-14}$ erg cm$^{-2}$ s$^{-1}$  (e) Absorption corrected, in units of $\rm 10^{42} \ erg \ s^{-1}$. (f) The value was frozen in the model fitting, therefore it was excluded in the analyses of the $\Gamma$ parameter variability of \autoref{sec:Xray_var}. (g) Values not provided by the authors.}
\end{table*}

\subsection{Historical Optical Spectra}

To access the historical presence or absence of \halpha BEL and the
galaxy optical classification, we reviewed all papers that report optical spectra of the galaxy. As most of these spectra are not available to allow inter-calibration and measurement of the BELs flux variability, we restrict ourselves to a binary classification of the  \halpha BEL, i.e., whether the component was undoubtedly detected (D) or not-detected (N-D) by the
authors. We do so by searching for a detection/non-detection claim in the
original text and by visually inspecting the published spectra. In
the middle panel of \autoref{fig:L_x_time}, we show the \halpha BEL detection history. In \autoref{table:optical}, we show the observation dates, the references and the Seyfert type at the time.

In summary: the first spectra of the galaxy in the '70s \citep{Ward_80,Veron_80,Shuder_80,Durret_88} detected only \halpha broad components, with no counterpart in \hbeta, resulting in its original classification as a Seyfert 1.9 galaxy. Throughout mid '80s to the mid '90s several authors have reported that the BEL \halpha was not detectable anymore \citep{Busko_90,Allen_99,Marquez_98,Veilleux_01}, changing its classification to Seyfert 2 Type. However the broad component was detected again from the end of the '90s to the beginning of the 2000s \citep{Gilli_00,Garcia-Lorenzo_01,Stoklasov_2009}. Later, \citet{Trippe_08} performed an 18 months long monitoring campaign from the beginning of 2006 to mid 2007 that revealed the \halpha BEL component was missing again, throughout the entire campaign. More recent observations have shown that the galaxy has regained the broad \halpha component \citep{Dopita_15,Schnorr-Muller_16,Mingozzi_19}, and for the first time, an \hbeta BEL was unambiguously detected, in late 2018, by \citet{Guolo_Pereira_2021}, therefore changing its classification to a 1.8 Type Seyfert galaxy.

\begin{table}
\begin{center}
    \caption{Historical broad \halpha line and Seyfert Type classification for NGC~2992.}
\begin{tabular}{|c|c|c|}
\hline
Obs. Date$^{(\rm a)}$ & H$\alpha$ BEL$^{(\rm b)}$ & Reference                      \\ \hline
07/1977        & D           & \citet{Ward_80}                             \\ 
< 1978         & D           & \citet{Veron_80}                           \\ 
01/1979        & D           & \citet{Shuder_80}                           \\ 
05/1979        & D           & \citet{Durret_88}                          \\ 
03/1985        & N-D          & \citet{Busko_90}                             \\ 
03/1991        & N-D          & \citet{Marquez_98}                          \\ 
02/1994        & N-D          & \citet{Veilleux_01}                           \\ 
03/1994        & N-D          & \citet{Allen_99}                              \\ 
04/1998        & D           & \citet{Garcia-Lorenzo_01}                   \\ 
01/1999        & D           & \citet{Gilli_00}                            \\ 
03/2001        & D           & \citet{Stoklasov_2009}                     \\ 
01/2006        & N-D          & \multirow{3}{*}{\citet{Trippe_08}}            \\  
12/2006        & N-D          &                                               \\  
06/2007        & N-D          &                                              \\ 
02/2014        & D           & \citet{Schnorr-Muller_16}                          \\ 
04/2014        & D           &   \citet{Dopita_15}                  \\ 
02/2015        & D           & \citet{Mingozzi_19}                      \\ 
02/2018        & D           &    \citet{Guolo_Pereira_2021}                                 \\
03/2021        & D           &    This Work                                 \\
04/2021        & D           &    This Work                               \\
\hline
\end{tabular}
\label{table:optical}

\end{center}

\footnotesize{(a) mm/yyyy. (b) Whether the \halpha
BEL was Detected (D) or Not-Detected (N-D).}
\end{table}

\subsection{Modern Optical Spectra}\label{sec:mod_spec}

For four of the six most recent published spectra of NGC~2992, we were able to apply methods of flux intercalibration that have been developed for reverberation mapping studies \citep{Peterson_93,van_Groningen_92,Peterson_95, Fausnaugh_2017} and measure the intrinsic flux variability of the BELs. 
The spectra are from: 

\begin{itemize}
    \item \citet[][]{Trippe_08} and taken in December of 2006 in the Cerro Tololo Inter-American Observatory; the spectrum was obtained by private communication with one of the authors. Spectra were observed using two different settings: one with a grating with a resolution of 4.3 \A (R = $\lambda/\Delta \lambda \sim$ 1000) to take blue spectra from approximately 3660-5440\AA, and the other with a resolution of 3.1 \AA~  (R $\sim$ 3000)  to take
    red spectra from approximately 5650-6970\AA;
    \item \citet[][]{Dopita_15} as part of the Siding Spring Southern Seyfert Spectroscopic Snapshot Survey (S7), taken in April of 2014 and publicly available at the S7 survey website. The spectra covers the waveband 3400-7100\AA~ with a high resolution of R = 7000 in the redder (5300-7100 \AA), and R = 3000 in the bluer (3400-5700 \AA) wavelengths;
    \item \citet{Mingozzi_19} as part of the Measuring Active Galactic Nuclei Under MUSE Microscope (MAGNUM) survey using the Multi-Unit Spectroscopic Explorer (MUSE) at the Very Large Telescope (VLT) and taken in February of 2015; the spectral data cube is public available at the VLT archive its spectral resolution goes from 1750 at 4650 \AA~ to 3750 at 9300 \AA;
    \item \citet{Guolo_Pereira_2021} taken using the Gemini Multi-Object Spectrographs (GMOS) at the Gemini South Telescope in February of 2018. The observed
    spectra covered a range from 4075 to 7285 \AA~ with and a spectral resolution of R $\sim$ 5000 at \halpha.
 
\end{itemize}

The analyses of Swift data signalled that the AGN flux was declining again to its fainter state, while this publication was being written, motivating us to obtain new optical spectra.  We proposed and obtained Director Discretionary Time (DDT) at the following Telescopes:

\begin{itemize}
    \item Southern Astrophysical Research Telescope (SOAR), using the SOAR Integral Field Spectrograph \citep[SIFS,][]{Lepine_2003} instrument, on the night of 08 March, 2021. Three 20 min exposures were made using a 700 l/mm grating, covering from 4500 to 7300\AA with a R $\sim$ 4200 spectral resolution;
    \item The Gemini South Telescope, using the Gemini Multi-Object Spectrographs (GMOS), on the night of 17 April, 2021. Two 10 min exposures were made using the same instrumental configuration and having the same spectral coverage and resolution as in  \citet[][]{Guolo_Pereira_2021}.

\end{itemize}
The standard reduction processes were performed in the two spectra data cubes: bias subtraction, flatfielding, trimming, wavelength calibration, telluric emission subtraction, relative flux calibration, the building of the data cubes, and finally the combination of the distinct exposures.

The inter-calibration, aperture/seeing corrections, and the process for emission line fitting of the six spectra will be presented in \autoref{sec:mod_optical}. It is important to note the large number of data points collected: 56 in the X-rays (49 with $N_{\rm H}$ and $\Gamma$ values) and 20 in the optical; which allow us to have a quasi-contemporaneous X-ray measurement for each optical spectrum, within less than one year for most, and less than two years for all of them, see \autoref{fig:L_x_time}.

\subsection{Estimates of the Black Hole Mass and Luminosity Distance}
\label{sec:BH_mass}

Precise black hole mass and luminosity distance measurements are essential for the energetic calibrations and scaling to come.
Using stellar velocity dispersion measurements of $\sigstar= 158$ \kms, \citet{Woo_2002} estimated the black hole mass of NGC\,2992 to be
$5.2 \times 10^{7}$\msun by applying the \citet{Tremaine_2002}  \Mbh-\sigstar relation. The value obtained from the \citet{Gultekin_09} \Mbh-\sigstar relation is \Mbh= $4.8^{+3.9}_{-2.4} \times
10^7$\msun, using a bulge stellar velocity dispersion \sigstar= $158 \ \pm$ 13\kms \citep{Nelson_95}. From the normalised excess variance \sigrms in the 2-10~keV band and assuming the  \sigrms-\Mbh correlation from \citet{Ponti_12}, \citet{Marinucci_20} reported a
 \Mbh$=3.0^{+5.5}_{-1.5}\times10^{7}$\msun. Furthermore, we also estimate the SMBH mass using the emission line measurements, to be described in \autoref{sec:fit}, by using single-epoch (SE) \Mbh estimates. We applied \citet{bonta2020} M$_{\textrm{SE}}(\sigma_{H\beta},L_{H\beta})$ relation using all six modern spectra (\autoref{sec:mod_spec}).
 We corrected the spectra for extinction effects using the measured Balmer decrements and assuming \citet{Calzetti_2000} extinction law. We obtained the following values for \Mbh: $1.3^{+3.8}_{-1.0}\times10^{7}$\msun, $3.5^{+9.1}_{-2.5}\times10^{7}$\msun, $2.4^{+3.2}_{-1.4}\times10^{7}$\msun, and
 $2.2^{+2.3}_{-1.1}\times10^{7}$\msun
 respectively for 2014, 2015, 2018 and 04/2021 spectra. As the \hbeta BEL was not detected in the 2006 and 03/2021 spectra, we use the \halpha BEL, assuming a Balmer decrement of 9 (see \autoref{sec:disc_opt}). We obtained, respectively, $1.2^{+1.5}_{-0.7}\times10^{7}$\msun and $2.5^{+2.7}_{-1.3}\times10^{7}$\msun. In \autoref{table:BH_mass} we summarise the estimated values, although the distinct methods results in distinct values, they are all within the $1-6\times10^{7}$\msun range.

 \citet{Theureau_2007} found a recession velocity of 2657\kms for NGC~2992 (already corrected for peculiar motions), which assuming a cosmology with $H_0$ = 70\kms Mpc$^{-1}$, translates to a luminosity distance to NGC~2992 equals to 38~Mpc.

\begin{table*}
\caption{Black hole mass estimates for NGC2992. Uncertainties are at 68\% (1 $\sigma$) confidence level.}
\begin{tabular}{|c|c|c|}
\hline$ \rm{M_{BH}}$                                                                                 & Method                                                                                                                                      & References                                                                                                                                 \\ \hline
$5.2\times10^{7}$ \msun                                                        & \begin{tabular}[c]{@{}c@{}}\Mbh-\sigstar \\  \citep{Tremaine_2002}\end{tabular}             & \begin{tabular}[c]{@{}c@{}}\citet{Woo_2002}\\ ($\sigstar= 158$ \kms)\end{tabular}                         \\ \hline
\begin{tabular}[c]{@{}c@{}}$4.8^{+3.9}_{-2.4} \times 10^7$\msun\end{tabular} & \begin{tabular}[c]{@{}c@{}}\Mbh-\sigstar \\ \citep{Gultekin_09}\end{tabular}             & \begin{tabular}[c]{@{}c@{}}\citet{Nelson_95}\\ (\sigstar= $158 \ \pm$ 13\kms)\end{tabular} \\ \hline
$3.0^{+5.5}_{-1.5}\times10^{7}$\msun                                           & \begin{tabular}[c]{@{}c@{}}\sigrms-\Mbh \\  \citep{Ponti_12}\end{tabular}                 & \citet{Marinucci_20}                                                                                                     \\ \hline
$1.2^{+1.5}_{-0.7}\times10^{7}$\msun                                           & \multirow{6}{*}{\begin{tabular}[c]{@{}c@{}}M$_{\textrm{SE}}(\sigma_{H\beta},L_{H\beta}) $ \\ \citep{bonta2020}\end{tabular}} & \begin{tabular}[c]{@{}c@{}}This Work \\ (2006 spectra)\end{tabular}                                                                        \\
$1.3^{+3.8}_{-1.0}\times10^{7}$\msun                                           &                                                                                                                                             & \begin{tabular}[c]{@{}c@{}}This Work\\  (2014 spectra)\end{tabular}                                                                        \\
$3.5^{+9.1}_{-2.5}\times10^{7}$\msun                                           &                                                                                                                                             & \begin{tabular}[c]{@{}c@{}}This Work \\ (2015 spectra)\end{tabular}                                                                        \\
$2.4^{+3.2}_{-1.4}\times10^{7}$\msun                                           &                                                                                                                                             & \begin{tabular}[c]{@{}c@{}}This Work \\ (2018 spectra)\end{tabular}                                                                        \\
$2.5^{+2.7}_{-1.3}\times10^{7}$\msun                                           &                                                                                                                                             & \begin{tabular}[c]{@{}c@{}}This Work \\ (03/2021 spectra)\end{tabular}                                                                     \\

$2.2^{+2.3}_{-1.1}\times10^{7}$\msun &               & \begin{tabular}[c]{@{}c@{}}This Work \\ (04/2021 spectra)\end{tabular}  
\\ 
\hline                                                       

\end{tabular}
\label{table:BH_mass}
\end{table*}
\section{Data Analyses}
\label{sec:analyses}

\subsection{X-ray Variability}
\label{sec:Xray_var}

\begin{figure}
    \centering
    \includegraphics[width=\columnwidth]{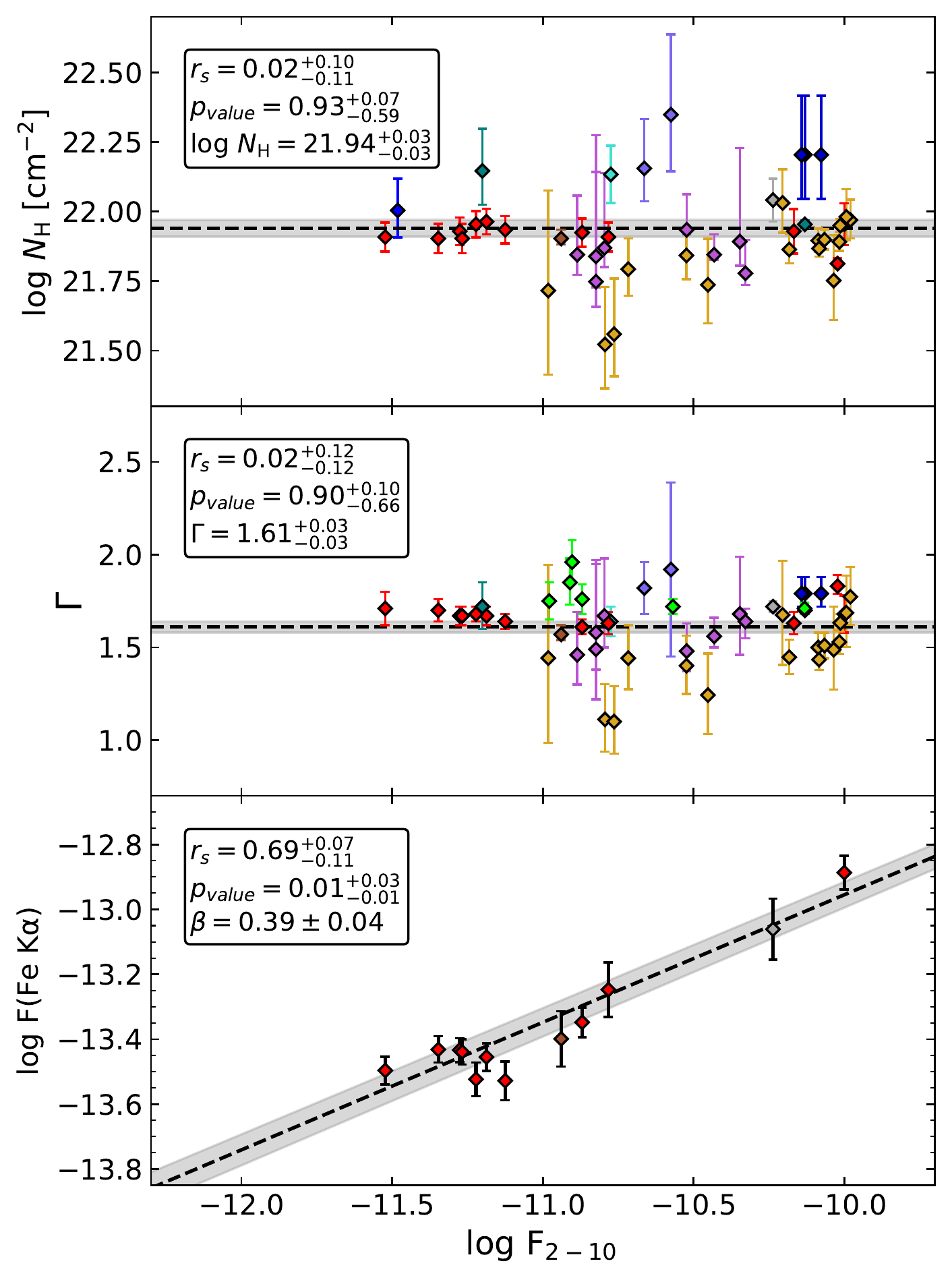}
    \caption{Best-fitting parameters N$_H$ (top panel) and $\Gamma$ (middle panel) for an absorbed power-law continuum model as function of the 2-10\,keV Flux. The left top boxes show the Spearman rank correlation coefficient ($r_s$) for the two variables and its corresponding $p_{\rm value}$, as well as the best-fitting constant values. Fe K$\alpha$ flux as function as function of the 2-10\,keV flux. The left top box show the Spearman rank correlation coefficient ($r_s$) for the two variables and its corresponding $p_{\rm value}$, $\beta$ is the slope of the best fitting line (see \autoref{eq:feka}).} 
    \label{fig:pars}
\end{figure}

In the scenario where the CL events are caused by variable absorption, two natural consequences are an anti-correlation between \Nh~and \Fx and a constant intrinsic (absorption corrected) luminosity. In the top panel of \autoref{fig:pars} we show ${\rm log}$ \Nh~as function of \Fx: no clear anti-correlation can be seen. The Spearman rank correlation coefficient ($r_s$) for the two variables and its corespondent $p_{\rm value}$ are  $r_s=0.02^{+0.10}_{-0.11}$ and $p_{\rm value}=0.93^{+0.07}_{-0.59}$, while the best-fitting constant value is \Nh = $21.93^{+0.03}_{-0.03}$, where the errors correspond to 90\% confidence level measured by performing a thousand Monte Carlo iterations (see details in appendix \ref{app1}).

We can therefore conclude that although there were some variations in the column density it is not anti-correlated with the X-ray flux as seen in variable absorption CL-AGNs \citep[e.g.][]{Puccetti_07,Bianchi_2009}. A further confirmation that the absorption is not the main factor responsible for the behaviour of the source can be seen by the intrinsic (i.e. absorption corrected) rest-frame luminosity (L$_{\rm 2-10}$). The measured L$_{\rm 2-10}$ values,
for each observation in which a $\Nh$ value is provided, are presented in \autoref{table:xray}. The values range from $4.9\times10^{41}$\ergs to $\sim$ $2.0\times10^{43}$\ergs. 
The L$_{\rm 2-10}$ light curve is shown in the bottom panel of \autoref{fig:L_x_time}, and it has the same behaviour as  F$_{\rm 2-10}$ confirming that the variations are intrinsic and not due to changes in absorption. 

The Spearman rank correlation test shows that there is no monotonic correlation between the photon index \G
and the X-rays flux (or luminosity) for NGC~2992 (middle panel of \autoref{fig:pars}). The values found were  $r_s=0.02^{+0.12}_{-0.12}$ and $p_{\rm value}=0.90^{+0.10}_{-0.66}$.  The existence of a correlation between luminosity and photon index in AGNs is in fact debatable, and several contradictory results have been presented in the last years. While \citet{Dai_2004} reported the existence of a positive correlation, other works found none \citep[e.g.][]{Winter_2009} or a negative correlation \citep[e.g.][]{Corral_2011}. A `V' shape for the \G versus \LEdd relation for NGC~2992 was claimed by \citet{liu_2019}; the authors interpreted this as a change in accretion mode, similar to the ones that occurs in stellar-mass X-ray binaries, e.g. \citet{Qiao_2013}. We argue here that their finding is due to selection effects, as the authors only used data from 2003 to 2013. When the entire historical data of the source is shown (\autoref{fig:pars}), this effect is not seen. In contrast, there is a positive correlation between \Fx and the flux of the Fe K$\alpha$ line, with $r_{s} = 0.69^{+0.07}_{-0.11}$ and p$_{\rm value} = 0.01^{+0.03}_{-0.01}$. Fitting the data by assuming a relation of the type:

\begin{equation}
    \label{eq:feka}
    {\rm log \ F(Fe \ K\alpha) = \alpha + \beta \ log \ \Fx}  
\end{equation}

\noindent
we obtained a slope of $\beta = 0.39\pm0.04$. As the iron K$\alpha$ line is created by reprocessing of the primary X-ray
continuum, a tight correlation between the flux of the
line and that of the continuum is expected.
For large samples of AGNs \citet{Ricci_2014}, found $\beta = 0.89\pm0.04$, while \citet{Shu_10} found $\beta = 0.86\pm0.01$. However, NGC~2992 seems to be one of a few objects where this correlation is found in a single object, though with a lower slope than the ones found in the general AGN population.

\subsection{Modern Optical Spectra Analyses}
\label{sec:mod_optical}

In this section, we explore the flux variability of the BELs using four of the six most recent spectra, namely from 12/2006, 04/2014, 02/2015, 02/2018, 03/2021, 04/2021 (see \autoref{table:optical}).  Uncertainties derived from any optical observation are at 68\% (1 $\sigma$) confidence level.

\subsubsection{Absolute calibration of the spectra}
\label{sec:abs}
Even under the best conditions, which are not
often realized, flux calibration of ground-based spectrophotometry is no better than $\approx$5\%,
which is insufficient for detailed comparisons of variable spectra. Thus the standard technique of flux calibration, through comparison with stars of known spectral energy distribution, is not good enough for the study of AGN variability. Instead, we use the fluxes of the narrow emission lines known to be non-variable on time scales of tens of years in most AGN. Consequently, the bright narrow emission lines can be adopted as internal calibrators for scaling AGN spectra \citep{Peterson_93}. We assume that the flux of the \oiii$\lambda$5007 line remains constant during the interval covered by these spectra.

We extracted the spectra using virtual apertures from the original data, making them as similar as possible to each other in the data cubes, as shown in \autoref{table:mod_spectra}. The scaling of the extracted spectra was carried out using a refinement of the method of
\citet{van_Groningen_1992}, implemented as a Python package ({\tt mapspec}) by \citet{Fausnaugh_2017}. This method allows us to obtain a homogeneous set of spectra with the same wavelength solution, same spectral resolution (at $\sim$ 6300\AA), and the same $\oiii\lambda$5007 flux value.

\begin{table*}
\caption{Modern Optical Spectra. Point-Source Scale Factor ($\varphi$) is measured using \citet{Peterson_95} algorithm, the errors in $\varphi$ represents the uncertainty in the centring of the virtual apertures. Uncertainties are at 68\% (1 $\sigma$) confidence level.}
\begin{tabular}{ccccccc}
\hline
\begin{tabular}[c]{@{}c@{}}Obs. Date\\ (mm/yyyy)\end{tabular} &
  Aperture &
  \begin{tabular}[c]{@{}c@{}}Point-Source\\ Scale Factor \\ $\varphi$ \end{tabular} &
  \begin{tabular}[c]{@{}c@{}}F(H$\alpha$)\\  (10$^{-13} \ {\rm erg \ cm^{-2} \ s^{-1}}$) \\ \end{tabular} &
  \begin{tabular}[c]{@{}c@{}}FWHM(H$\alpha$)\\ (\kms) \\\end{tabular} &
  \begin{tabular}[c]{@{}c@{}}F(H$\beta$)\\ (10$^{-13} \ {\rm erg  \ cm^{-2} \  s^{-1}}$) \\ \end{tabular} &
   F(H$\alpha$)/F(H$\beta$) \\ \hline
12/2006 & Slit (width = 2.0\arcsec)   & 0.734 $\pm$ 0.063 & $\leq$ 4.4    & 2733 $\pm$ 55 &--              & --               \\
04/2014 & Circular (radius = 2.8\arcsec) & 0.972 $\pm$ 0.025 & 27.5 $\pm$ 1.4 & 2002 $\pm$ 28 & 3.6 $\pm$ 2.6      & 7.6 $\pm$ 2.8             \\
02/2015 & Circular (radius = 3.0\arcsec) & 1.000 & 33.8 $\pm$ 1.5 & 2313 $\pm$ 17 & 3.3 $\pm$ 2.2   & 10.1 $\pm$ 3.3 \\
02/2018 & Circular (radius = 3.0\arcsec) & 0.973 $\pm$ 0.013 & 43.8 $\pm$ 2.2 & 2016 $\pm$ 13 & 4.9 $\pm$ 1.4 & 8.8 $\pm$ 1.7 \\ 
03/2021 & Circular (radius = 3.0\arcsec) & 1.000 $\pm$ 0.012 & 25.4 $\pm$ 2.0 & 2209 $\pm$ 15 & $\leq$ 3.1 & $\geq$ 8.2  \\
04/2021 & Circular (radius = 2.0\arcsec) & 0.862 $\pm$ 0.009 & 18.5 $\pm$ 0.5 & 2225 $\pm$ 5 & 1.7 $\pm$ 0.1 & 10.3 $\pm$ 0.9  \\\hline
\end{tabular}
\label{table:mod_spectra}

\end{table*}

\subsubsection{Stellar component subtraction and emission-line fitting}
\label{sec:fit}

We subtracted the host galaxy stellar continuum fitted using the \textsc{starlight} \citep{Cid_Fernandes_2005} full spectra fitting code, using a base of 45 Simple Stellar Population (SSP) spectra, with 15 ages (from 1 Myr to 13 Gyr) and 3 metallicities (0.2, 1.0 and 2.5\zsun). The emission line fitting was performed using the \textsc{IFSCube} package \citep{ifscube}. The code employs Sequential Least Square Programming and allows multiple components with physically motivated constraints. In order to measure the BEL fluxes, the narrow emission components must be separated from the broad components. This process is not straightforward due to the fact the BELs are blended with the narrow lines, for both \halpha and \hbeta.

\begin{figure}
    \centering
    \includegraphics[width=\columnwidth]{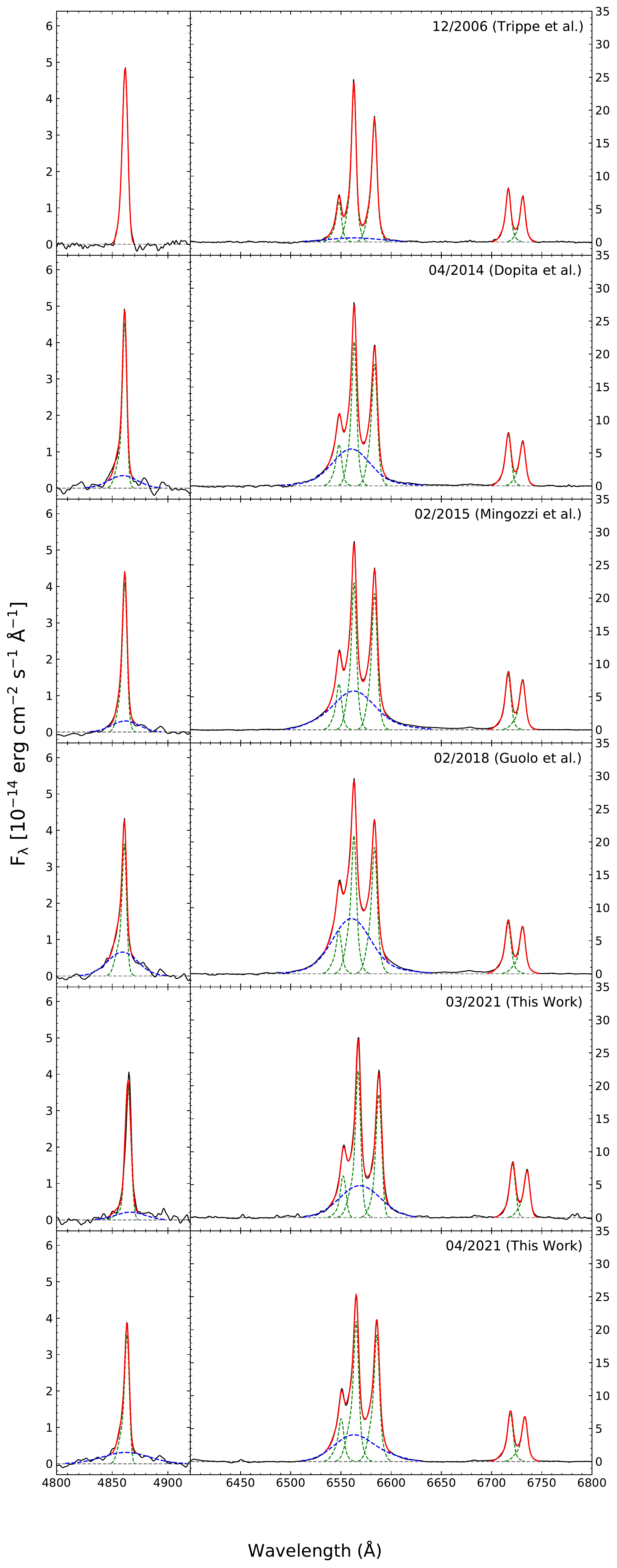}
    \caption{Fit of the emission lines for the modern optical spectra. From top to bottom: 12/2006 \citep{Trippe_08}, 04/2014 \citep{Dopita_15}, 02/2015 \citep{Mingozzi_19}, 02/2018 \citep[][]{Guolo_Pereira_2021}, and the new spectra from March and April 2021. Scaled observed spectra (black), narrow components (green), broad components (blue) and total model (red) are shown. Left panels show the \hbeta lines while right ones show the \halpha, \nii and \sii lines.}
    \label{fig:BEL_fit}
\end{figure}

The narrow line subtraction was performed in two steps. First we fitted the narrow and broad \halpha and \hbeta simultaneously. For the narrow lines,
we created a template comprised of two Gaussian components
that were based on the $\sii\lambda$6716 profile. This template
was used to fit the \nii and narrow \halpha and \hbeta lines by adjusting
its flux scaling factor only. The fits to the broad lines were
subtracted from the spectra, isolating the narrow emission; we then re-fitted the narrow lines assuming the \ion{H}{i} and \nii
lines have the same velocity dispersion and redshift and we
set the flux of $\nii\lambda$6583 line as 2.96 times that of the
$\nii\lambda$6548 line, in accordance with the ratio of their transition probabilities \citep{Osterbrock_2006}, while the \halpha
and $\nii\lambda$6583 fluxes were free parameters in the fit. Finally, we subtracted the resulting narrow lines from the original spectrum and re-fit only broad lines using three Gaussian components, in order to account for the known non-Gaussian profiles of the BEL. The errors were taken as the standard deviations of one hundred Monte Carlo iterations. The resulting fits for the four spectra are shown in \autoref{fig:BEL_fit}.

\subsubsection{Inter-calibration of the spectral data}\label{sec:disc_opt}

The fluxes measured in the scaled spectrum described in the previous sections were corrected for aperture and distinct seeing effects, because
while the BLR is effectively a point-like source, the NLR is an extended one. Consequently, the measured NLR flux depends on the size of the spectrograph’s entrance aperture and the observations seeing \citep[see][ for a detailed discussion]{Peterson_95}. In order to correct our fluxes for these effects, we determined a point-source correction factor ($\varphi$), which was measured using the procedure described in \citet{Peterson_95}.
The $\varphi$ value represents the differences in the amount of light from the NLR with respect to the amount of light from the BLR for each observation, it is normalised using the largest aperture ($\varphi \equiv 1.0$, for 04/2015 spectrum). Then the scaled inter-calibrated and aperture/seeing corrected BEL fluxes were measured as follows:

\begin{equation}
F(H\alpha) = \varphi \cdot F_{\lambda5007} \left [ \frac{F(H\alpha)}{F(\oiii\lambda5007)} \right ]_{\rm obs}
\end{equation}

\noindent
where F$_{\lambda5007}$ is the scaling absolute flux in the $\oiii\lambda$5007 line, described in \autoref{sec:abs}, and the value in brackets is the broad \halpha to $\oiii\lambda$5007 measured flux ratio for each spectrum as described in \autoref{sec:fit}. The same procedure was applied to \hbeta BEL. The BEL \halpha and \hbeta fluxes and the Balmer decrements (\halpha/\hbeta) are shown in \autoref{table:mod_spectra} and in the right panels of \autoref{fig:BEL_flux}. In order to consider a valid BEL detection we required a signal to noise ratio (S/N) greater than 5; if the BEL is not detected above this level we report the upper limit flux.
In the 12/2006 spectrum, the \halpha BEL is not unambiguously detected (which leads to its classification as a Seyfert 2), and if it is indeed present, it is very faint, with an upper limit flux of $4.4\times10^{-13}$\ergcms.
From 2014 to 2018 the broad \halpha flux increases from $\sim28\times10^{-13}$\ergcms to $\sim 44\times10^{-13}$\ergcms.
The new 2021 spectra, however, show a decline in the \halpha BEL flux.
In 03/2021 we measured a flux of $\sim 25\times10^{-13}$\ergcms, while in 04/2021 the flux was $\sim 18\times10^{-13}$\ergcms.
The same trend is present in the \hbeta flux, which evolves from undetected in 2006 to an increase from 2014 until 2018, and finally declining in 2021.
Within the uncertainties the Balmer decrements seems to be constant at a value of $\sim$ 9.

\subsection{The Link Between the X-ray Luminosity and the BELs}
\label{sec:link}
Comparing the middle and bottom panels of \autoref{fig:L_x_time}, one can promptly see a correlation between the intrinsic X-ray luminosity and the detection of the broad \halpha component.
The periods in which the broad component is detectable are the ones where L$_{2-10}$ shows the highest values; epochs in which the galaxy is classified as a Seyfert 2 are those when the closest available X-ray spectra show lower L$_{2-10}$ values.
Given that the X-ray and optical observations are not simultaneous, in order to quantify the correlation between the BELs and the X-ray luminosity we linearly interpolate the  L$_{2-10}$ curve (bottom panel of \autoref{fig:L_x_time}) and attribute a  L$_{2-10}$ value for each historical optical spectrum. In \autoref{fig:BLR_x_Edd} we show the interpolated L$_{2-10}$ value for each spectrum as function of the corresponding \halpha BEL detection.

\begin{figure}
    \centering
    \includegraphics[width=1.0\columnwidth]{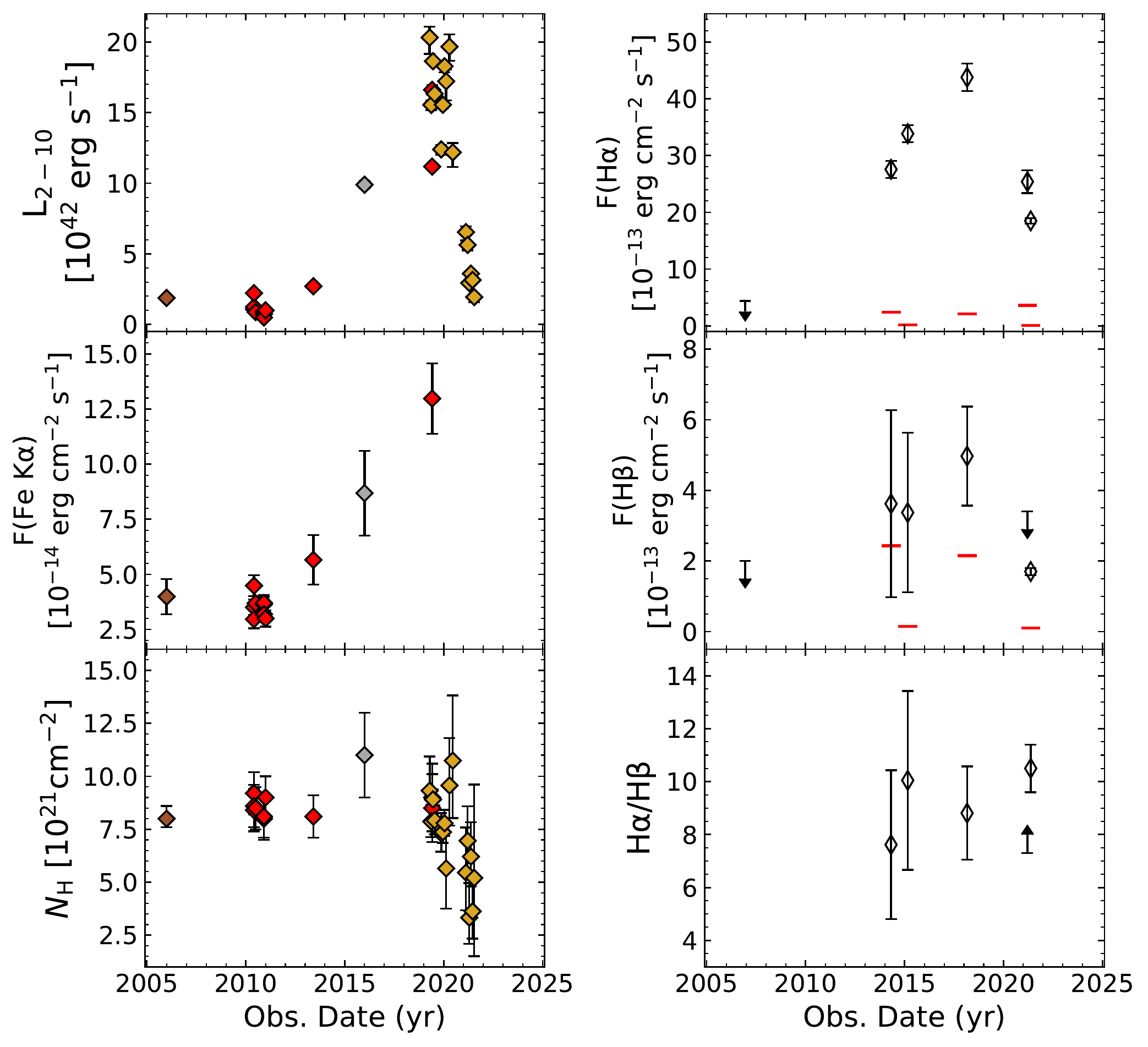}
    \caption{X-ray (left) and optical (right) properties from 2005 to 2020. From top to bottom in the X-ray panels: Intrinsic 2-10 keV Luminosity (L$_{2-10}$), narrow Fe K$\alpha$ line flux, and the column density (\Nh). From top to bottom in the optical panels: Flux of the \halpha BEL, the flux of the \hbeta BEL, and the Balmer decrement (\halpha/\hbeta). In the optical panels, the red lines represent a lower limit for the detection of the lines, being measured as the flux of a line with a S/N equals to 5. In the cases in which the emission lines are not detected above this S/N, the upper limits are shown.}
    \label{fig:BEL_flux}
\end{figure}

From \autoref{fig:BLR_x_Edd}, the link between the BEL and the X-ray luminosity is clear: at all epochs when an \halpha BEL is not detected, L$_{2-10}$ < $2.6\times10^{42}$ erg s$^{-1}$ (the X-ray luminosity of the bright Seyfert 2 spectra); conversely, when a Seyfert 1.x spectrum is observed, L$_{2-10}$ > $2.6\times10^{42}$ erg s$^{-1}$.
Moreover, from \autoref{fig:BEL_flux} (top panels) we can see that for the modern spectra (2006-2021) the \halpha BEL flux increases with the increase in the L$_{2-10}$.
From the same figures, we can see that the transitions between Type 1.9 to Type 1.8 (i.e., the clear detection of the \hbeta BEL) also seems to be synchronous to L$_{2-10}$.
It is possible that the oldest high Eddington Ratio optical observations did not detect the broad \hbeta due to its faintness or lower signal to noise ratio.

We can convert the transitional X-ray luminosity ($2.6\times10^{42}$ erg s$^{-1}$) into a Bolometric luminosity (\Lbol) and then into the Eddington ratio ($\lambda_{\rm Edd}$), as follows:

\begin{equation}\label{eq:lamb_edd1}
    \lambda_{\rm Edd} =\frac{L_{\rm Bol}}{L_{\rm Edd}} =  \frac{\left [ K_{\rm X}(L_{2-10})\pm \Delta K_{\rm X} \right ] \times L_{2-10}}{1.3\times10^{38} \ {\rm M}_{\rm BH}}
\end{equation}

\noindent 
where K$_{\rm X}$ is the bolometric correction and $\Delta$K$_{\rm X}$ is uncertainty in the bolometric correction (the intrinsic spread of the AGN population dominates this uncertainty). We use \citet{Duras_2020} correction, which assumes values of K$_{\rm X}$(L$_{2-10}$ = 2.6$\times 10^{42}$ erg s$^{-1}$) = 15.54 and $\Delta K_{\rm X} = 0.37~{\rm dex}$. Therefore the Eddington ratio at which the \halpha BEL disappears can be written as :

\begin{equation}\label{eq:lamb_edd2}
    \lambda_{\rm Edd} \approx 1.1^{+1.4}_{-0.7} \ \times \left ( \frac{3\times10^{7}\msun}{\Mbh} \right)  \ \%
\end{equation}

\noindent
where the term ${\rm (3\times10^{7}\msun/\Mbh)}$ is written as a systematic scaling factor to take into account the uncertainties in the black hole mass and the distinct values that can be considered (see \autoref{sec:BH_mass}).
Using the same prescription, we show in the top panel of \autoref{fig:FWHM_x_Edd} the clear correlation between the measured \halpha BEL fluxes and $\lambda_{\rm Edd}$. 
Assuming the above predicted transitional X-ray luminosity (2.6$\times 10^{42}$ erg s$^{-1}$) is correct, the AGN should be at its Seyfert 2 state as of June 2021 (see \autoref{table:xray}). Unfortunately, we were not able to get a new optical spectrum, given that the galaxy is not visible in the sky at the time of this publication. We will try to do a follow-up observation as soon as the galaxy becomes visible again, in November 2021, in order to confirm this hypothesis.

In summary, this analysis strongly supports the hypothesis that the CL phenomenon in NGC~2992 is caused by variations in the accretion rate and not due to changes in absorption nor due to TDEs (which we discard due to the several repeated CL events).

In the last several years, the idea that the orientation-based Unification Models alone cannot explain the variety of AGNs \citep[see][for a review of challenges to the Unification Model]{spinoglio_2019}, and that the specific accretion rate, as given by the Eddington ratio, is as essential as obscuration in the classification of the distinct AGN types, has been explored by several authors. For example, \citet[][]{Trump_2011} using a sample of 118 unobscured (\Nh~< 10$^{\rm 22}$ cm$^{\rm -2}$) AGNs show that the BELs are present only at the highest accretion rates ($\lambda_{\rm Edd} > $ 1\%). 

Several authors have proposed models that posit different accretion rates as a cause of the differences between observed AGNs \citep[see][for a review of the properties of low luminosity AGNs]{Ho_2008}. \citet{Elitzur_2009} suggest that the BLR and ``torus'' are inner (ionized) and outer (clumpy and dusty) parts of the same disk-driven wind and that this wind is no longer supported at low accretion rates \citep[][]{Nicastro_2000,Elitzur_2006,Nenkova_2008}. Based on this disk-wind BLR model, \citet{Elitzur_2014} proposed that the intrinsic spectral sequence S1\f S1.2/S1.5\f S1.8/S1.9\f S2\footnote{Referring here only to the ``true'' Type 2 Seyferts, meaning those which do not have BELs even in polarised light, therefore excluding those which have obscured Type 1 nucleus.} is a true evolutionary sequence, reflecting the evolution of the BLR structure with decreasing accretion rate onto the central black hole and that this evolution is regulated by the \Lbol/\Mbh$^{2/3}$ variable (which behaves like $\lambda_{\rm Edd}$). These models have been successfully used to explain the existence of ``true'' Type 2 Seyferts as shown by \citet{Marinucci_2012} and are recently being invoked to explain CL-AGNs. 

\subsection{Proposed Scenarios}

We have shown that the variability in NGC~2992 is intrinsically driven by changes in accretion rate, and not due to variable obscuration. In this scenario, the CL events (i.e., the disappearing and reappearing of the BELs) can be due to two distinct effects:

\begin{itemize}
    \item Dimming/brightening of the AGN continuum, which changes the supply of ionising photons available to excite the gas in the immediate vicinity of the black hole and therefore the BLR. In this case, variations in the BEL fluxes are due to changes in the ionisation state of the BLR clouds and not to change in its geometry or structure \citep{Storchi-Bergmann_2003,Schimoia_2012,Lamassa_15};
    \item The fading of the BLR structure itself as proposed by the disk-wind BLR models \citep{Elitzur_2006, Elitzur_2009}, in which a low accretion rate is not able to sustain the required cloud flow rate \citep{Nicastro_2000,Trump_2011, Elitzur_2014}, that makes up the BLR.
\end{itemize}

\begin{figure}
    \centering
    \includegraphics[width=0.95\columnwidth]{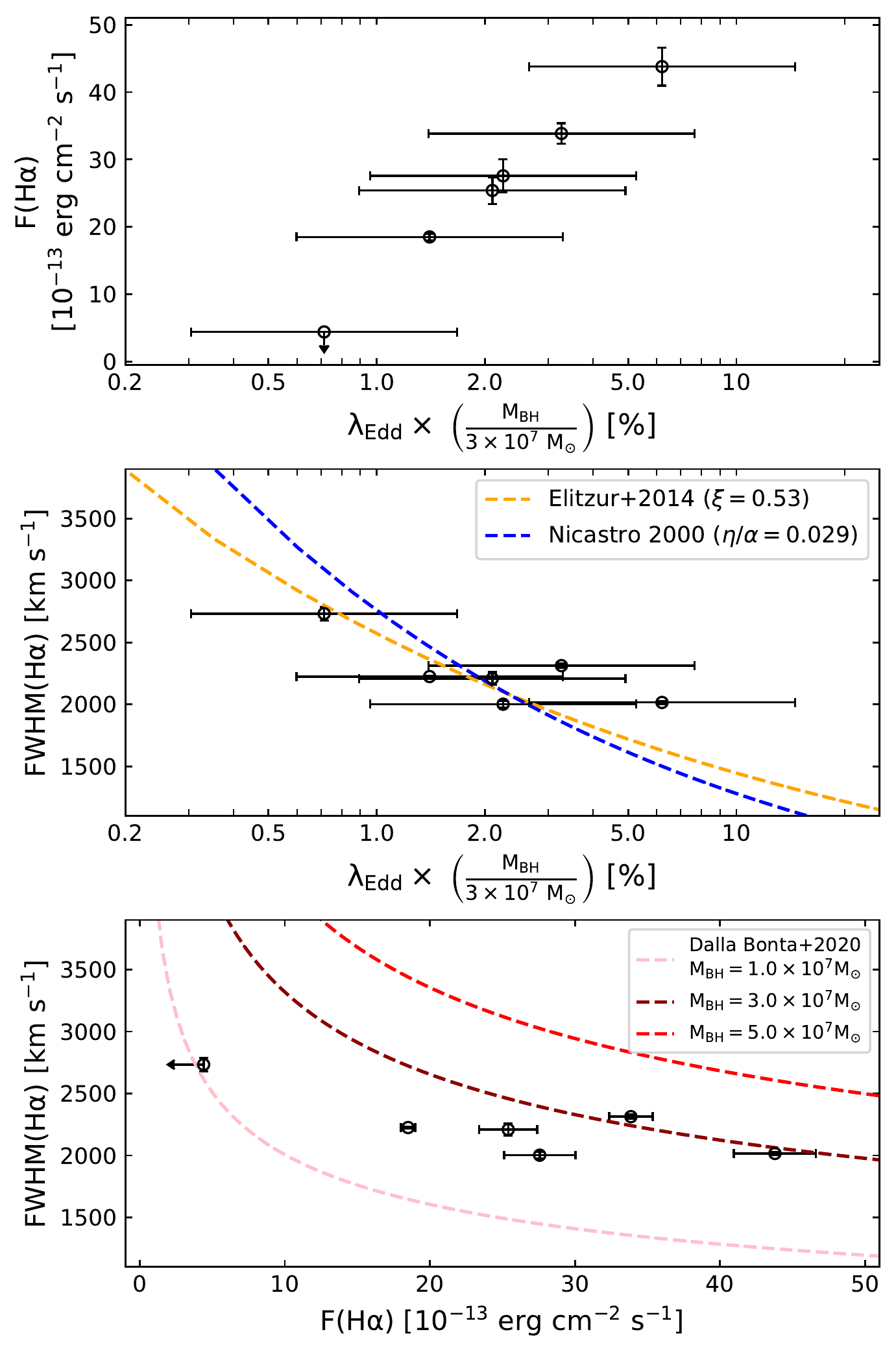}
    \caption{Top: flux of \halpha BEL as function of Eddington Ratio ($\lambda_{\rm Edd}$), measured using equations \ref{eq:lamb_edd1} and \ref{eq:lamb_edd2}, for each modern optical spectrum. Middle: FHWM of \halpha BEL as function of $\lambda_{\rm Edd}$ for each modern optical spectrum. Blue and orange dashed lines are, respectively, presents the best-fitting \citet{Nicastro_2000} and \citet{Elitzur_2014} BLR models (see \autoref{sec:link} and equations \ref{eq:nicastro} and \ref{eq:elitzur}). The term ${\rm (3\times10^{7}\msun/\Mbh)}$ is written as a systematic scaling factor to take into account the uncertainties in the black hole mass and the distinct values that can be considered (see \autoref{sec:BH_mass}). Bottom: FHWM of \halpha BEL as function of \halpha BEL flux for for each modern optical spectrum. The dashed lines represent the \citet{bonta2020} relation between the two parameters, assuming a constant Balmer decrement of 9 and \citet{Calzetti_2000} extinction law, for distinct supermassive black hole masses (\Mbh) in the range discussed in \autoref{sec:BH_mass}.}
    \label{fig:FWHM_x_Edd}
\end{figure}

The first scenario is supported in the case of the double-peaked  LINER/Seyfert 1 nucleus of NGC\,1097, for which \citet{Storchi-Bergmann_2003} and \citet{Schimoia_2012} found an inverse correlation between the width of the \halpha broad profile and the flux of the line, as well as for NGC 5548 by \citet{Peterson_2002} and for the CL-Quasar SDSS J015957.64+003310.5 by \citet{Lamassa_15}. The authors preferred explanation for this is the fact that, when the AGN is brightest, it ionises farther out in the BLR, reaching lower velocity clouds, and the profile becomes narrower. When the AGN is dimmer, it ionises only the closest regions, where the velocities are higher and the profile becomes wider, in agreement with a virialized BLR. In our data, there seems to be such a inverse correlation between FWHM(\halpha) and F(\halpha), in the modern optical spectra, as can be seen in the bottom panel of \autoref{fig:FWHM_x_Edd}. 

\begin{figure}
    \centering
    \includegraphics[width=0.8\columnwidth]{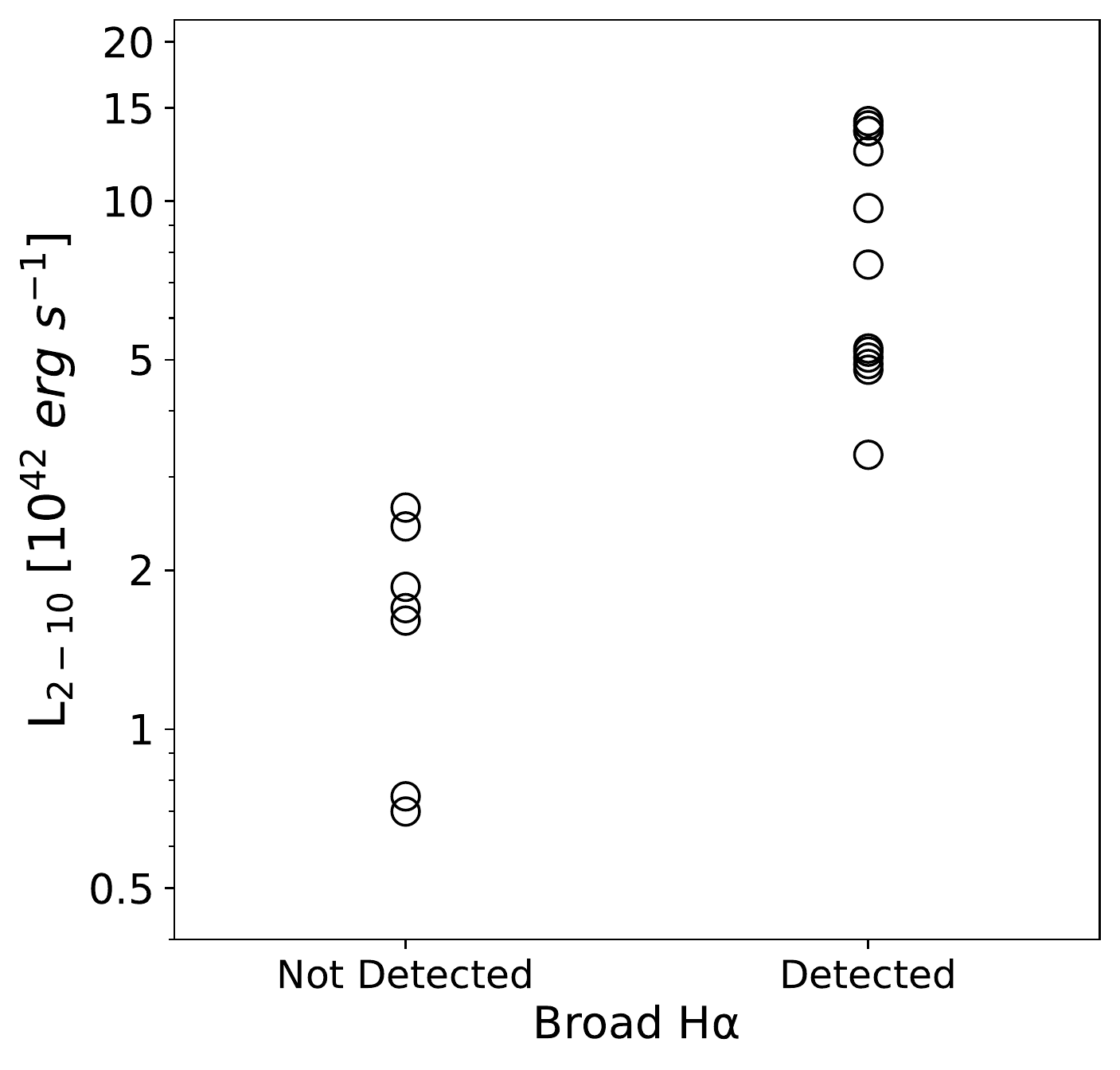}
    \caption{Interpolated intrinsic 2-10 keV x-ray luminosity (L$_{2-10}$) for each optical spectrum as function of the detection or non-detection of the \halpha BEL. Showing a clear boundary of L$_{2-10}$ $\approx 2.6\times10^{42}$ erg s$^{-1}$ for the disappearance of the broad \halpha component}
    \label{fig:BLR_x_Edd}
\end{figure}

However, in the second scenario, where the BLR is described as a disk-driven wind and not as clouds orbiting in Keplerian-like orbits, this inverse correlation is also predicted. Using distinct, prescriptions both \citet{Nicastro_2000} and \citet{Elitzur_2014} show that with increasing accretion rate ($\lambda_{\rm Edd}$) (which increases \halpha BEL flux) the FWHM in the BLR decreases. In the \citet{Nicastro_2000} model, the relation between the two parameters is written as:

\begin{equation}
    \label{eq:nicastro}
    {\rm log}(\lambda_{\rm Edd}) = -3 {\rm log}({\rm FWHM}) + 9.86 + {\rm log}(\eta/\alpha) 
\end{equation}

\noindent
where $\eta/\alpha$ is the ratio between the disk efficiency ($\eta$) and its viscosity ($\alpha$), while \citet{Elitzur_2014} use:

\begin{equation}
    \label{eq:elitzur}
    {\rm log}(\lambda_{\rm Edd}) = -4 {\rm log}({\rm FWHM}) + 3.60 + {\rm log}(\Mbh) + {\rm log}(\xi) 
\end{equation}

\noindent where $\xi$ is the ratio between the BLR radius ($r_{ BLR}$) and the dust sublimation radius ($R_{d}$) with $\xi < 1.0$.
The reader is referred to the original papers as well as to \citet{Nicastro_2018} for detailed explanation of the models and equations. 
In the middle panel of \autoref{fig:FWHM_x_Edd}, we show both models with best-fitting parameters ($\eta/\alpha = 0.029$ and $\xi = 0.53$) for the measured  \halpha BEL FWHM and $\lambda_{\rm Edd}$, measured from the interpolated L$_{2-10}$ (\autoref{fig:BLR_x_Edd}) using the above prescription (equations \ref{eq:lamb_edd1} and \ref{eq:lamb_edd2}).

While we have successfully shown that the CL events in NGC~2992 are driven by changes in the accretion rate, we could not determine the true nature of the BEL variability, mainly because, as discussed, we still do not have a full understanding of the formation and evolution of the BLRs in AGNs. In the next section we discuss the overall state-of-art modelling of the innermost structure of active galaxies as well as a proposed technique to distinguish between the two above proposed scenarios.

\section{Results and Discussion}
\label{sec:discussion}

In both the scenarios discussed in the previous section, quick ($\sim$ 10 years between the maximum and the minimum activity) and large (a factor of $\sim$ 40 in L$_{2-10}$) variations in the accretion rate are necessary. However, the standard \citet{Shakura_73} accretion disk model does not support such variations, given that for AGN scales (where the model still predicts the existence of a geometrically thin disk, H $\ll$ R ) the viscous time scale of the disk is in the order of $10^4$ to $10^5$ years, greatly exceeding the timescales over which CL transitions are known to occur. This incompatibility was noticed long ago \citep[e.g.][]{Koratkar_1999,Antonucci_2013}, but just recently some modification or alternatives to the application of \citet{Shakura_73} model to AGNs have been proposed in light of the increasing number of the these CL sources been discovered.

For example, \citet{Dexter_2018} propose that AGN accretion discs are vertically supported by magnetic pressure which makes it geometrically thick (H/R between 0.1 and 1.0) at all luminosities. \citet{Jiang_2020} show, for a $5\times10^{8}$\msun mass black hole, using three dimensional radiation magneto-hydrodynamic simulations, that under this assumption the Rosseland mean opacity is expected to be larger than the electron scattering value, and the iron opacity bump then causes the disk to be convectively unstable. This results in strong fluctuations in surface density and heating of the disk. The opacity drops with increasing temperature and convection is suppressed. The disk cools down and the whole process repeats. This causes strong oscillations of the disk scale height and luminosity variations by more than a factor of $\approx$ 3 -- 6 over timescales of few years. They argue that since the iron opacity bump will move to different locations of the disk for black holes with different masses and accretion rates, this is a physical mechanism that can explain the accretion rate variability of AGN with a wide range of amplitudes over a time scale of years to decades.

The Eddington ratio in which the BEL transitions occurs in NGC 2992 contains the critical value at which there is a state transition between a radiatively inefficient accretion flow (RIAF) and a thin accretion disk \citep[$\lambda_{\rm Edd} \  \sim$ 1\%; e.g.][]{Xie_2012,Ho_2008}. The similarity between these values suggests that NGC 2992 is operating at the threshold mass accretion rate between the two accretion modes. Indeed, \citet{Sniegowska_2020} proposed an explanation for the sources displaying multiple quasi-periodic CL phenomena according to which they would be operating at a few per cent of the Eddington limit. They argue that the outbursts are caused by the radiation pressure instability operating in the narrow ring between the standard \citep{Shakura_73} thin pressure dominated outer disk and the hot optically thin inner RIAF \citep{Yuan_2014}. The corresponding limit cycle is responsible for periodic outbursts, and the timescales are therefore much shorter than the standard viscous timescale due to the narrowness of the unstable radial zone. 

Even with the possibility of such rapid variability in the accretion rate (given some of these alternatives to the classical accretion disk) the cause of the type transition remains uncertain, dimming of the ionising source or fading of the BLR structure itself. By using radiative transfer Monte Carlo simulations to test the multiple causes of the CL phenomenon, \citet{Marin_2017} argues that the differences between these two scenarios can only be observable in polarised light: in the ionising source dimming scenario the polarisation properties between the high and low flux state should be the same, while that for the BLR fading a decrease in the degree of polarisation and a rotation in the polarization angle are expected in the low flux state. We had access to a spectropolarimetric observation of NGC~2992
obtained in 2006 (Robinson et al., in prep), when the galaxy was at a dim state (see \autoref{fig:L_x_time}). It shows no signs of BEL in the polarimetric spectrum (as well as in normal flux). In order to test \citet{Marin_2017} predictions we would need another polarimetric spectrum, with the AGN at its bright state. As this data is not available for now, the exact intrinsic cause of the CL-AGN events in NGC~2992 is still an open question. We join these authors in advocating for systematic polarimetric observations of CL-AGN in order to fully understand their true nature.
In this sense, follow up observations (polarimetric or not) of NGC~2992 in the next decades will continue to help constrain the physics of AGNs. Furthermore, this source joins several others in the literature that show that the accretion rate is a fundamental variable for any model that tries to unify the AGN zoology, and that viewing angle or obscuration effects alone cannot fully explain differences seen in SMBH activity in galaxies.

\section{Conclusions}
\label{sec:conclusion}

We have analysed historical X-ray and optical observations of the Seyfert galaxy NGC~2992, from 1978 to 2021 -- as well as presented new X-ray and optical spectra -- focusing on its CL events. The main conclusions are:

\begin{itemize}
    
    \item The source presents large intrinsic X-ray luminosity variability ranging from $4.9\times10^{41}$\ergs up to $2.0\times10^{43}$\ergs;
    \item Several transitions between type 2 Seyfert and intermediate-types are documented, with the appearance and disappearance of the \halpha BEL occurring on timescales of several years;
    \item We ruled out TDEs or variable obscuration as causes of the type transitions. We show that the flux of the \halpha BEL is directly correlated with the 2-10 keV X-ray
    luminosity: the \halpha BEL seems to disappear at L$_{2-10}$ values lower than $2.6\times10^{42}$ erg s$^{-1}$. This value translates to an Eddington ratio of $1.1^{+1.4}_{-0.7}$\%, if we assume \Mbh = $3\times10^{7}$\msun.
    \item We find a correlation between the narrow Fe K$\alpha$ line flux and $\lambda_{\rm Edd}$, and an anti-correlation between full width at half maximum of the H$\alpha$ BEL and $\lambda_{\rm Edd}$, as predicted by theoretical work by \citet{Nicastro_2000} and \citet[][]{Elitzur_2012}.
    \item Two possible scenarios for type transitions are still open: either the dimming (brightening) of the AGN continuum luminosity, which reduces (increases) the supply of ionising photons available to excite the gas in the immediate vicinity of the black hole \citep{Lamassa_15} or the disappearance of the BLR structure itself occurs as the low accretion efficiency is not able to sustain the required cloud flow rate in a disk-wind BLR model \citep{Elitzur_2006, Trump_2011, Elitzur_2014}. Multi-epoch polarimetric observations may distinguish between the two scenarios. 
    \item This study supports the idea that the accretion rate is a fundamental variable in determining the observed AGN type, and therefore should be included in a more complete model to unify the AGN zoology.
    
\end{itemize}

\section*{Acknowlegments}
We would like to thank all those authors and observatories that made public their data making this work possible, specially Michael Crenshaw for providing the \citet{Trippe_08} spectrum and Adam Thomas for helping us access the S7 survey \citep{Dopita_15} data.  We also thank Gemini's and SOAR's staff, in special Luciano Fraga, for their rapid response on accepting and observing our DDT proposals.
MGP also would like to thanks Michael Fausnaugh for making {\tt mapspec} public available as well as for his comments and suggestions on the paper.

Based  in part  on observations obtained at the Southern Astrophysical Research (SOAR) telescope, which is a joint project of the Minist\'{e}rio da Ci\^{e}ncia, Tecnologia e Inova\c{c}\~{o}es (MCTI/LNA) do Brasil, the US National Science Foundation’s NOIRLab, the University of North Carolina at Chapel Hill (UNC), and Michigan State University (MSU). Based on observations obtained at the international Gemini Observatory, a program of NSF’s NOIRLab, which is managed by the Association of Universities for Research in Astronomy (AURA) under a cooperative agreement with the National Science Foundation on behalf of the Gemini Observatory partnership: the National Science Foundation (United States), National Research Council (Canada), Agencia Nacional de Investigaci\'{o}n y Desarrollo (Chile), Ministerio de Ciencia, Tecnolog\'{i}a e Innovaci\'{o}n (Argentina), Minist\'{e}rio da Ci\^{e}ncia, Tecnologia, Inova\c{c}\~{o}es e Comunica\c{c}\~{o}es (Brazil), and Korea Astronomy and Space Science Institute (Republic of Korea). We acknowledge the use of public data from the Swift data archive.

\section*{DATA AVAILABILITY}
The GMOS, MUSE and Swift data are public available at the Gemini, ESO, and HEASARC archive web pages, respectively. The other data underlying this article will be shared on reasonable request to the corresponding author.


\bibliographystyle{mnras}
\interlinepenalty=10000
\bibliography{references}



\appendix

\section{Propagation of Asymmetric Uncertainties}
\label{app1}

When the underlying likelihood function of a parameter is not symmetric, the error distribution will also be asymmetric. Different origins of asymmetry are discussed  in detail by \citet[][]{Barlow_2003,Barlow_2003b} and \citet{Barlow_04}. In high-energy astrophysics it is related to the low photon counting. In this case, the estimated parameters are usually reported in a form like $\hat{x}^{+u_{\rm r}}_{-u_{\rm l}}$. Where $\hat{x}$ is the best-fitting parameter (minimum $\chi^2$) and $u_{\rm l}$ and $u_{\rm r}$ are the draw where the $\Delta\chi^2$ curve reaches some value, for classical 1$\sigma$ (68\%) confidence level $\Delta\chi^2 = 1.0$, however the 90\% confidence level ($\Delta\chi^2 =2.7$) is the most employed in X-ray astronomy \citep{Avni_76,Lampton_76,Yaqoob_98}.

Usually, one needs to calculate a function of one or more quantities, and propagate its uncertainties. In most circumstances, well known ``Error Propagation'' formulas are adequate. But there are some assumptions and approximations under the Error Propagation, and if they fail, the method is invalid. These assumptions can be summarised as: the errors are Normally distributed (Gaussian errors) and they are relatively small. 
For asymmetric errors in astronomy any of these assumptions can be made; if the asymmetry is really small, rounding/averaging the negative and the positive errors into a single value and making the distribution symmetric does not change the result much. But when the values are not appropriate to do that, researchers often tend to put two error values into quadrature separately and combine them in the result. Actually this is not a valid method due to violation of the Central Limit Theorem \citep{Barlow_2003b}. Eventually more sophisticated methods are needed to use: we present here a Python-based Monte Carlo solution for this problem, based in a combination of \citet{Barlow_04} and \citet{Possolo_19} derivations. 

During this research we faced the problem that even with numerous of uncertainties propagation tools publicly-available (e.g. \href{https://docs.astropy.org/en/stable/uncertainty/}{\tt astropy.uncertainty},
\href{https://pythonhosted.org/uncertainties/}{\tt uncertainties}) none of them are able to handle
asymmetric uncertainties as we needed, due to the assumptions stated above. We therefore created the \href{https://github.com/muryelgp/asymmetric\_uncertainties}{\tt asymmetric\_uncertainties} Python package, the main idea of the code is to generate
a random sample from the likelihood function and use this sample in other calculations
\citep{Possolo_19}. The code assumes a ``Variable Width Gaussian'' likelihood, which, as shown by
\citet{Barlow_04}, reproduces very well the underlying likelihood of low counting
measurements, and, in contrast to, for example, a ``Generalised Poisson'' distribution, has a unique and analytical solution for given values of $\hat{x}$, $u_{\rm l}$ and $u_{\rm r}$ at a given confidence level ($\Delta\chi^2$). From this
likelihood the code generates a random sample of data, that can be passed as parameters
for any type of function. Simple operations are deal internally by the code and the more
sophisticated ones can be easy performed by the user, for example: L$_{\rm 2-10}$ values in
\autoref{table:xray} whose uncertainties were propagated from \G and \Nh~uncertainties; as
well as the statistical tests (r$_s$ and p$_{\rm value}$) in \autoref{fig:pars}.
The resulting sample is then fitted by the code returning the nominal value and the
propagated uncertainties. The code is publicly-available at
\href{https://github.com/muryelgp/asymmetric_uncertainties}{github}\footnote{https://github.com/muryelgp/asymmetric\_uncertainties} where examples in how to use the code are also presented.


\bsp	
\label{lastpage}
\end{document}